\documentclass[preprint,showpacs,preprintnumbers,amsmath,amssymb]{revtex4-1}

\usepackage{bm}
\usepackage{amsmath}
\usepackage{amsfonts}
\usepackage{graphicx}

\newcommand{\nn}{\nonumber}
\newcommand{\bs}[1]{\boldsymbol{#1}}

\usepackage{color}

\newcommand{\zs}{\sigma}

\newcommand{\ze}{\varepsilon}
\newcommand{\zb}{\beta}
\newcommand{\zg}{\gamma}

\newcommand{\zm}{\mu}

\newcommand{\zr}{\rho}
\newcommand{\za}{\alpha}

\newcommand{\zf}{\varphi}

\newcommand{\zh}{\eta}

\newcommand{\zp}{\phi}

\newcommand{\zO}{\Omega}
\newcommand{\zS}{\Sigma}

\newcommand{\dif}{\; \textrm d }

\newcommand{\dpp}[2]{\frac{  \partial #1 }{  \partial #2   } }
\newcommand{\dpt}[2]{\frac{  \dif #1 }{  \dif #2   } }

\begin{document}
\title{Wigner model for quantum transport in graphene.}
\author{O. Morandi$^{1}$, F. Schuerrer$^{1}$}
\affiliation{$^1$ Institute of Theoretical and Computational Physics, TU Graz,
Petersgasse 16, 8010 Graz, Austria}
\begin{abstract}
The single graphene layer is a novel material consisting of a flat monolayer of carbon atoms packed in a two-dimensional honeycomb-lattice, in which the electron dynamics is governed by the Dirac equation. A pseudo-spin phase-space approach based on the Wigner-Weyl formalism is used to describe the transport of electrons in graphene including quantum effects. Our full-quantum mechanical representation of the particles reveals itself to be particularly close to the classical description of the particle motion. We analyze the Klein tunneling and the correction to the total current in graphene induced by this phenomenon. The equations of motion are analytically investigated and some numerical tests are presented. The temporal evolution of the electron-hole pairs in the presence of an external electric field and a rigid potential step is investigated. The connection of our formalism with the Barry-phase approach is also discussed.
\end{abstract}
\date{\today}
\maketitle

\section{Introduction}
Graphene can be considered as one of the most splendid functional materials. This has been proved by the quick science respond to the novel fascinating experiments performed by A. K. Geim and K. S. Novoselov \cite{Geim07}.
Graphene represents a single layer of $sp^2$-bonded carbon atoms, which are densely packed in form of a benzene ring structure. This ideal planar structure has been used to describe properties of many carbon-based materials including graphite (that can be viewed as a large number of superposed graphene sheets).
This new, strictly two-dimensional material displays unusual electronic properties arising from the bi-conically shaped form of the Fermi surfaces near the Brillouin zone corners (Dirac points). In a quite wide range of energy, electrons and holes propagate as massless Fermions and their behavior reproduces the physics of quantum electrodynamics but at the much smaller energy scale of the solid state physics.

For example, a typical superconductivity phenomenon, like the Josephson effect, have analogs in the p-n junction. In \cite{Beenakker_08_b} the particle transport trough an interface between a normal and a supercoductor material (N-S) is compared with the analogous inter-band tunneling in a p-n graphene junction. It is shown that for excitation energies, which are small compared to the superconducting gap, the Dirac Hamiltonian of an p-n junction displays the same excitation spectrum as an N-S junction.

New experiments preformed in graphene-based materials showed the evidence of the simultaneous occurrence of relativistic-like and superconducting transport that opens the possibility to study the implication of the relativistic-like electron behavior in the solid state. In \cite{Heersche_07} Heersche et al. studied the supercurrent flowing through a simple device constituted by two superconducting electrodes on the top of a carbon monolayer. They observed the occurrence of some new properties displayed by massless particles in graphene as for example the integer quantum Hall effect and the Aharonov-Bohm effect \cite{Zhang_09,Morozov_06}. Moreover, a large Rashba splitting (corresponding to an energy shift of 225 meV) of the $\pi$ states in an epitaxial graphene layer on a Ni substrate has been reported in \cite{Dedkov_08}.

In particular, evidence has mounted that the scattering of electrons near the Dirac point in graphene-superconductor junctions differs from the analogous Andreev scattering process in normal metals and that the quasi-ballistic transport in graphene sheets is only weakly affected by external sources of disorder (defects or impurities). In fact, Dirac fermions are quite immune to the localization effects and it has been observed experimentally that electrons can propagate without scattering over distances of the order of micrometers \cite{Novoselov_04}.
Furthermore, in high-density low-temperature regimes, the mobility is roughly density- and temperature-independent. Although the interactions with the underlying substrate are largely responsible for the relaxation properties of particles in graphene (possible sources of scattering include adsorbents and defects in the graphene lattice, ionized impurities in the silicon oxide substrate, surface charge traps, interfacial phonons and substrate ripples \cite{Chen_08}), the exact nature of the scattering that limits the mobility of electrons in graphene devices remains unclear.

Graphene is a semiconductor, whose band gap is exactly zero and the velocity of the charge carriers is over a wide range of energy independent of the momentum. Graphene is expected to be in a low-conductivity state when the Fermi energy approaches the Dirac point where the density of states vanishes. A gate voltage can, however, modulate the density of states in graphene and switch between the low-conductivity state at the Dirac point and the high-conductivity states elsewhere.  The charge mobility in graphene layers attains large values that cannot be reached in conventional semiconductors (mobility of the order of $10^5 \,\textrm{cm}^2 \textrm{V}^{-1} \textrm{s}^{-1}$ have been recently measured \cite{Bolotin_08}).
Because of this high electronic mobility and the capability of being tuned from $p$-type to $n$-type doping by the application of a gate voltage, graphene is an interesting candidate towards possible applications in carbon-based electronics devices. In particular, some applications are already devised by various groups, for example in designing electronic building blocks \cite{Lemme_07,Williams_07} or spin injection devices \cite{Tombros_07}.

Moreover, when the Fermi level approaches the Dirac point, the density of states vanishes and it is expected that also the conductivity becomes strictly zero. On the contrary, the theoretical prediction of Fradkin, given in \cite{Fradkin_86}, concerning the presence of a residual minimal charge conductivity, was confirmed by experiments. The main reason of this phenomenon concerns the difficulty of localizing Dirac-like particles in a single band. The possibility to perform easily band-to-band transitions, provided by the gap-less
Dirac-like form of the Hamiltonian, reveals that the particles can travel over long distances (or penetrate a potential barrier) without creating a reflected component by converting itself in a electron-hole excitation.
Because of the strong similarity with relativistic quantum mechanics, the tunneling of an electron through an p-n graphene junction, where conduction-like states are converted into hole-like states (and viceversa) is denoted as Klein tunneling. It represents the tunneling of a particle into the Dirac sea of antiparticles (represented by the almost filled hole band).
In several recent experiments, this unusual coupling of electron- and hole-like dynamics have been investigated \cite{Beenakker_08}. Klein tunneling gives rise to some unusual behavior of the charge transport when the Fermi level approaches the Dirac point, where the valence and conduction bands meet. In particular, an unusual interesting transport phenomenon of relativistic-like particles concerns the normal incidence of a particle-antiparticle beam on a square potential barrier. When the incidence angle becomes equal to zero, the barrier becomes completely transparent (Klein paradox). This result is characteristic for the Dirac-like dispersion relation of the Hamiltonian and contrasts the electron transport in normal (nonrelativistic) devices, where the band-to-band transmission probability is always smaller than one.

In solid state physics, we are typically interested in macroscopic phenomena, which are slowly varying in time and smooth in space apart from variations on the atomic scales. The language used to describe electron transport is derived from the semi-classical picture of the dynamics where the electrons respond to external fields like point particles. There have been overwhelming evidences that such a simple picture cannot give complete account of  first-order effects in the fields.

The development of efficient quantum computational methods is thus a crucial aspect in the study of new devices where quantum-mechanical effects play a dominant role.
Different approaches based on the density matrix, non-equilibrium Green's functions, and the Wigner function have been proposed to achieve a full quantum mechanical description of the electron transport \cite{Haug_Jauho_book}. Among them, the Wigner-function formalism is the one that bears the closest similarities to the classical Boltzmann equation, so that this formalism can be considered as a natural choice to derive quantum corrections to the classical phase-space motion. Furthermore, a phase-space approach may appear more intuitive compared with the more abstract density matrix and Green's function formalism. The phase-space formulation of quantum mechanics offers a framework in which quantum phenomena can be described with a classical language and the question of the quantum-classical correspondence can be directly investigated \cite{morandi_JPA_10}.

For these reasons, an approach where both the kinetic characteristic of the particles and the pseudospin degree of freedom are described in a full-quantum mechanical framework, seems to be a promising approach to shed light on these particularities of graphene.
The close similarity between the classical mechanics and a quantum kinetic framework, which characterizes the Wigner single-band formalism, is generally lost when we address the many-band dynamics. In fact, a straightforward extension of the standard definition of the Wigner function leads to very complicated multi-band systems, where a one-to-one relationship between band and distribution function cannot be found. In general, it is not possible to define a quasi-distribution function associated to a single kind of particles (holes or electrons) and whose marginal distributions (for example the integral with respect the momentum) represents some expectation values of such particles.

In this contribution, we therefore present a Wigner-like multiband formalism and study the effect of Klein tunneling. In sec. \ref{sec mod} the derivation of the quasi-diagonal equations of motion is presented and our approach is compared with some preexisting methods. In sec. \ref{sec unif ele} we discuss the particles motion in the presence of a uniform electric field. Particular emphasis is given on the close similarity between the description of the Klein tunneling provided by our method and the classical particle transport. The numerical difficulties for a direct solution of the transport equations are discussed in sec. \ref{sec eff mod} and an asymptotic approach is proposed. Finally, in sec. \ref{sec num sol full quant} we study the particle motion in the presence of a rigid barrier.

\section{Wigner formalism for the quantum transport in graphene}\label{sec mod}

The atomic structure of graphene is characterized by two types of bonds and exhibits the so-called planar sp$^2$ hybridization.  The $\zs$ bonds are strong covalent bonds responsible for most of the binding energy and for the elastic properties of the graphene sheet. However, since the upper (lower) bound of the $\zs$ ($\zs^*$) band is quite faraway the Fermi energy (more than 4 eV and 8 eV at the $\Gamma$ point for the $\zs$ and the $\zs^*$ orbital, respectively), bonding and anti-bonding $\zs$ bands can be safely neglected when addressing the electronic properties of graphene. The half-filled $\pi$ bands are responsible for the charge transport.
The first who studied the graphene band structure was P. R. Wallace in 1946 by using a tight binding approach \cite{Wallace_47}. Subsequently, more refined models were derived, providing a reliable theoretical basis for the description of the electronic properties of this material (an exhaustive bibliography concerning this models can be found in \cite{Neto_09}).
The Hamiltonian  \cite{Beenakker_08,Novikov_07}
\begin{eqnarray}
   \widehat{\mathcal{H}} &=&  \widehat{\mathcal{H}}_{0} + \zs_0 U(\mathbf{r}) \; ,\\
   \widehat{\mathcal{H}}_{0} &=& - i \;  v_F\hbar \;  \bs{\zs} \cdot \nabla_{\mathbf{r}}  =    v_F\hbar  \left(\begin{array}{cc}
0 & -i\dpp{}{x}  - \dpp{}{y} \\
  -i\dpp{}{x}  + \dpp{}{y} & 0
                                    \end{array}\right) \; , \label{hamlit oper graph}
\end{eqnarray}
reproduces the spectrum of an electron-hole pair in a graphene sheet lying in the $x-y$ plane, in the presence of an external potential $U(\mathbf{r})$. Here, $v_F$ is the Fermi velocity, $\bs{\zs}= \left(\zs_x,\zs_y,\zs_z \right) $ denotes the Pauli matrices vector and $\zs_0$  the identity $2\times2$ matrix.
The valence and conduction bands are usually denoted as pseudo-spin components of the particle.

From a technological point of view, the direct integration of a graphene sheet in a device could cause some disadvantages.
They are mainly related to the absence of an energy gap between the particles and holes at the Fermi surfaces that prevents the electrons to be  electrostatically confined in graphene.
If compared with other open gap carbon-based structures, as for example carbon nano-ribbons (where band gaps of nearly 1 eV are observed), the absence of a gap allows high current flows also in the off state. This strongly limits the application of a graphene sheet as a suitable channel in a carbon-based FET. For this reason, we derive our evolution model for a more general Hamiltonian than Eq \eqref{hamlit oper graph}, containing an energy gap $\Delta$ at $\mathbf{p}=0$:
\begin{eqnarray}
   \widehat{\mathcal{H}} &=&  \widehat{\mathcal{H}}_{\Delta} + \zs_0 U(\mathbf{r}) \label{hamlit oper graph open gap} \; ,\\
   \widehat{\mathcal{H}}_{\Delta} &=& \widehat{\mathcal{H}}_{0} +\Delta  \; \zs_0\; .
\end{eqnarray}

We establish the particle equation of motion in the quantum kinetic formalism by defining a suitable multi-component Wigner function. From a technical point of view, one of the main purposes of our approach is to describe the particle evolution by a set of Wigner functions $f_{ij}(\mathbf{r},\mathbf{p})$ in such way that each function $f_{ij}$ is the Wigner transform of a mixture of electronic states belonging only to the $i$-th and $j$-th band. This ambitious goal would require the diagonalization of the pseudo-spinorial Hamiltonian $\widehat{\mathcal{H}}$ of Eq. \eqref{hamlit oper graph open gap} in the momentum as well as in the position space. However, this is in general impossible, due to the non-commutativity of these operators. In the following, we propose a procedure that tries to define a set of basis states that diagonalize ``as much as possible" $\widehat{\mathcal{H}}$. For that reason, the corresponding set of Wigner functions will be denoted as a ``quasi-diagonal" representation.

We study the electron-hole pair system by means of the Weyl quantization procedure. For the sake of completeness, we recall briefly the basic mathematical tools used in the Weyl formulation. Given a differential operator $\widehat{\mathcal{A}}$ (defined on a suitable Hilbert space $\mathbb{H}$) and a function $h$, the Weyl map  $\mathcal{W}\left[\mathcal{A}\right] (h)= \widehat{\mathcal{A}} h $, is defined as
\begin{eqnarray}\label{def simbolo}
\left( \hat{\mathcal{A}} h \right)(\mathbf{x}) &=&  \frac{1}{\left(2\pi\hbar \right)^d} \int
 \mathcal{A} \left( \frac{\mathbf{x}+\mathbf{y}}{2}, \mathbf{p} \right)\,h (\mathbf{y})\, e^{\frac{i}{\hbar}(\mathbf{x}-\mathbf{y})\cdot
 \mathbf{p}}\dif \mathbf{y}\dif \mathbf{p}\; .\nn
\end{eqnarray}
It establishes a unique correspondence between $\widehat{\mathcal{A}}$ and a function $\mathcal{A}(\mathbf{r},\mathbf{p})$ which is denoted as the symbol of the operator. Here $d$ is the dimension of the position and momentum space. 
In the framework of the Weyl quantization procedure, a mixed state is defined by the density operator
$$
\widehat{\mathcal{S}}  [h] = \int \zr  (\mathbf{x},\mathbf{x}') h(\mathbf{x}') \dif \mathbf{x}'
$$
whose kernel is the density matrix. The Weyl symbol $\mathcal{S} = \mathcal{W}^{-1} \left[\hat{\mathcal{S}} \right]$
is obtained by applying the inverse of the Weyl transformation (Wigner transformation) to the function $\zr (x,x')$ \cite{Zachos_book}:
\begin{eqnarray}
\mathcal{S} (\mathbf{r},\mathbf{p}) =    \int  \zr  \left(\mathbf{r}+\frac{\bs{\zh}}{2},\mathbf{r} -\frac{\bs{\zh}}{2}\right) e^{-\frac{i}{ {\hbar}} \mathbf{p}\cdot \bs{\zh}}\dif \bs{\zh}\; . \label{wigner transf}
\end{eqnarray}
The von Neumann equation
\begin{eqnarray}
i \hbar \dpp{\widehat{\mathcal{S}}   }{t}   &=&   \left[\widehat{\mathcal{H}} ,\widehat{\mathcal{S}}   \right]\label{eq neu 1}
\end{eqnarray}
gives the evolution of the density operator and expresses the evolution of the system in operational form. By using the Weyl operator, this equation can be mapped into an evolution equation defined in the phase plane $\mathbf{r}-\mathbf{p}$. The symbol associated to the graphene Hamiltonian given in Eq. \eqref{hamlit oper graph open gap} is  $\mathcal{H}\left(\mathbf{r},\mathbf{p}\right)\equiv\mathcal{W}^{-1} \left[\widehat{\mathcal{H}}\right] =  \mathcal{H}_\Delta(\mathbf{p}) +   \zs_0 U(\mathbf{r})$ where $\mathcal{H}_\Delta = v_F \bs{\zs} \cdot \mathbf{p}+\Delta\;\zs_0  $ (in this simple case, the usual quantization $ - i \hbar \nabla \rightarrow \mathbf{p} $ holds true).
We consider the density operator $\widehat{\mathcal{S}'}  \equiv \widehat{\Theta} \; \widehat{\mathcal{S}}\; \widehat{\Theta}^\dag  $ where $\widehat{\Theta}  \left(\mathbf{r},\nabla_{\mathbf{r}} \right)$ is a unitary $2\times 2$ matrix operator and the superscript $\dag$ denotes transposition and conjugation. A convenient quantum kinetic description of the electron-hole pair motion can be obtained if we exploit the link of $ \widehat{\Theta} $ with the symbol $\Theta \left(\mathbf{r},\mathbf{p}\right) \equiv  \mathcal{W}^{-1} \left[\widehat{\Theta} \right] $. In particular, we require that $\Theta \left(\mathbf{r},\mathbf{p}\right)$ diagonalizes the Hamiltonian $\mathcal{H}\left(\mathbf{r},\mathbf{p}\right)$ \emph{locally} in the position and in the momentum space. We have
\begin{eqnarray}
   {\Theta}   {\mathcal{H}}_\Delta {\Theta}^\dag   &=&   \Lambda \label{diag H}\\
   \Theta (\mathbf{p})  &=& \frac{1}{\sqrt{2 E }}  \left(
                                \begin{array}{cc}
                                \sqrt{E+ \Delta }    & \displaystyle e^{-i\theta_{\mathbf{p}}}   \sqrt{E-  \Delta }   \\
                                 \displaystyle  e^{ i\theta_{\mathbf{p}}}   \sqrt{E-  \Delta }    &  - \sqrt{E+ \Delta }    \\
                                \end{array}
                              \right) \label{theta symb}
\end{eqnarray}
where $\Lambda(\mathbf{p}) = \zs_z E (\mathbf{p}) $, the relativistic-like spectrum of the graphene sheet $E=\sqrt{v_F^2 |\mathbf{p}|^2 +\Delta^2 }$, and $ e^{ i\theta_{\mathbf{p}}} = \frac{p_x+ip_y}{\sqrt{p_x^2+p_y^2}}$.
Equation (\ref{eq neu 1}) transforms to
\begin{eqnarray}
i \hbar \dpp{\widehat{\mathcal{S}'}  }{t}   &=&   \left[\widehat{\mathcal{H}'},\widehat{\mathcal{S}'} \right] \; , \label{mot Neum primo}
\end{eqnarray}
where $\widehat{\mathcal{H}'} \equiv \widehat{\Theta} \;  \widehat{\mathcal{H}} \; \widehat{\Theta}^\dag $. By applying the operator $\mathcal{W}^{-1}$ to Eq. (\ref{mot Neum primo}), we obtain the final equation of motion for the symbol $\mathcal{S}' = \mathcal{W}^{-1} \left[ \widehat{\mathcal{S}'}\right]$ in the phase-space $\left(\mathbf{r},\mathbf{p}\right)$ (details of calculations are given in Appendix \ref{app for H Hp}):
\begin{eqnarray}
i \hbar \dpp{ \mathcal{S}'  }{t}   &=&   \left[\mathcal{U}'+ \Lambda(\mathbf{p}) , \mathcal{S}' \right]_\star \; ,  \label{mot Neum sec}
\end{eqnarray}
where the brackets denote commutation $\left[\mathcal{A} , \mathcal{B} \right]_\star= \mathcal{A}\star\mathcal{B} -\mathcal{B} \star \mathcal{A}$. The star-Moyal product $\star $ is defined as
\begin{eqnarray}
\mathcal{A}\star \mathcal{B}   \equiv \mathcal{A} \; e^{\frac{i\hbar}{2} \left(\overleftarrow{\nabla_\mathbf{r}} \cdot \overrightarrow{\nabla_\mathbf{p}}- \overleftarrow{\nabla_\mathbf{p}} \cdot \overrightarrow{\nabla_\mathbf{r}}\right) } \; \mathcal{B}\; , \label{star prod}
\end{eqnarray}
where the arrows indicate on which operator the gradients act. The symbol $\mathcal{U}'\left(\mathbf{r},\mathbf{p}\right) $ is given by \begin{eqnarray}
\mathcal{U}'\left(\mathbf{r},\mathbf{p}\right)  &=& \Theta \star     U \left(\mathbf{r} \right)    \star  \Theta^\dag  \label{rela fra sym}
\end{eqnarray}
and writes explicitly as
\begin{eqnarray*}
\mathcal{U}'(\mathbf{r},\mathbf{p})& =&   \frac{1}{\left(2\pi\right)^2 }\int   {\Theta}  \left(  \mathbf{p}+\frac{ \hbar}{2}  \boldsymbol{\mu}  \right)    {\Theta}^\dag   \left(  \mathbf{p} - \frac{ \hbar}{2} \boldsymbol{\mu}    \right)  U (\mathbf{r}'  ) e^{i (\mathbf{r}-\mathbf{r}')\cdot \boldsymbol{\mu} }   \dif  \boldsymbol{\mu} \dif \mathbf{r}' \; .
\end{eqnarray*}
Equation \eqref{mot Neum sec} is given in terms of the Moyal commutator and defines implicitly a non-local evolution operator for the matrix-Wigner function $\mathcal{S}'$. It requires the evaluation of infinite-order derivatives with respect to the variables $\mathbf{r}$ and $\mathbf{p}$. The commutators appearing in Eq. \eqref{mot Neum sec} can be written in integral form as
\begin{eqnarray}
\left[\Lambda  , \mathcal{S}' \right]_\star
&=&\frac{1}{\left(2\pi\right)^{ 2} }  \int \left[\Lambda  \left(  \mathbf{p} +\frac{ \hbar}{2} \boldsymbol{\mu} \right)   \mathcal{S}'  \left( \mathbf{r}'  ,\mathbf{p}  \right)  - \mathcal{S}' \left( \mathbf{r}'  ,\mathbf{p}  \right)  \Lambda \left( \mathbf{p} -\frac{ \hbar}{2} \boldsymbol{\mu} \right)  \right]  e^{i (\mathbf{r}-\mathbf{r}')\cdot \boldsymbol{\mu}  }   \dif  \boldsymbol{\mu}     \dif \mathbf{r}' \nn\\\label{[Lam S]} \\
%
\left[\mathcal{U}' , \mathcal{S}' \right]_\star
&=&\frac{1}{\left(2\pi\right)^{4} }  \int \left[\mathcal{U}' \left( \mathbf{r}-\frac{ \hbar}{2}   \boldsymbol{\zh} ,\mathbf{p} +\frac{ \hbar}{2} \boldsymbol{\mu} \right)   \mathcal{S}' \left( \mathbf{r}' ,\mathbf{p}' \right)  - \mathcal{S}' \left( \mathbf{r}' ,\mathbf{p}' \right)  \mathcal{U}'\left( \mathbf{r}+\frac{ \hbar}{2}   \boldsymbol{\zh} ,\mathbf{p} -\frac{ \hbar}{2} \boldsymbol{\mu} \right)  \right] \nn \\ && \hspace{8cm }\times \; e^{i (\mathbf{r}-\mathbf{r}')\cdot \boldsymbol{\mu} + i (\mathbf{p}-\mathbf{p}')\cdot \boldsymbol{\zh}}   \dif  \boldsymbol{\mu} \dif \mathbf{r}'  \dif  \boldsymbol{\zh} \dif \mathbf{p}' \;. \nn \\ \label{[U S]}
\end{eqnarray}
The commutator of Eq. \eqref{[Lam S]} describes the free motion of the electron-hole pairs in the upper and lower conically shaped energy surfaces $\Sigma^\pm$. One of the principal aims of the diagonalization procedure of Eq. \eqref{diag H} was to derive an equation of motion, where the free motion is described in terms of the evolution of two non-interacting particle populations. This is achieved since $\Lambda$ is a diagonal matrix. The free evolution of the particles $f^+$ ($f^-$) belonging to the upper (lower) part of the spectrum is described by
\begin{eqnarray}
\dpp{f^{\pm}}{t} &=& \pm \frac{1}{\left(2\pi\right)^{ 2} }  \int \left[E \left(  \mathbf{p} +\frac{ \hbar}{2} \boldsymbol{\mu} \right)   -   E \left( \mathbf{p} -\frac{ \hbar}{2} \boldsymbol{\mu} \right)  \right]  f^{\pm} \left( \mathbf{r}'  ,\mathbf{p}  \right) e^{i (\mathbf{r}-\mathbf{r}')\cdot \boldsymbol{\mu}  }   \dif  \boldsymbol{\mu}   \dif \mathbf{r}' \; ,  \label{sing band app}
\end{eqnarray}
where we defined the components of the matrix $\mathcal{S}'$ as
\begin{eqnarray}
\mathcal{S}'\equiv (2\pi\hbar)^2 \left(
                 \begin{array}{cc}
                  f^+(\mathbf{r},\mathbf{p})  & f^i (\mathbf{r},\mathbf{p})\\
                  \overline{f^i}(\mathbf{r},\mathbf{p}) & f^- (\mathbf{r},\mathbf{p})\\
                 \end{array}
               \right) \; . \label{comp Sp}
\end{eqnarray}
These equations describe the free quantum mechanical motion in the band structure defined semi-classically by the function $E(\mathbf{k} ) =\sqrt{v_F^2 \hbar^2 \left| \mathbf{k} \right|^2+\Delta^2 }$  and generalizes the mass term present in the parabolic band approximation. It should be noted that our procedure is derived in a full quantum mechanical context, without invoking the usual generalization of the semi-classical motion to the quantum mechanical one, where the substitution $ \mathbf{k} \rightarrow - i \nabla_{\textbf{r}} $ in the semiclassical expression of the energy spectrum $E(\mathbf{k})$ is assumed.
As expected from a physical point of view,  the coupling between the bands arises from the presence of an external field $U(\mathbf{r})$ which perturbs the periodic crystal potential. This is described by Eq. \eqref{[U S]}.

\begin{figure}[!t]
\begin{center}
a)\includegraphics[width=0.450\textwidth,height=0.37\textwidth]{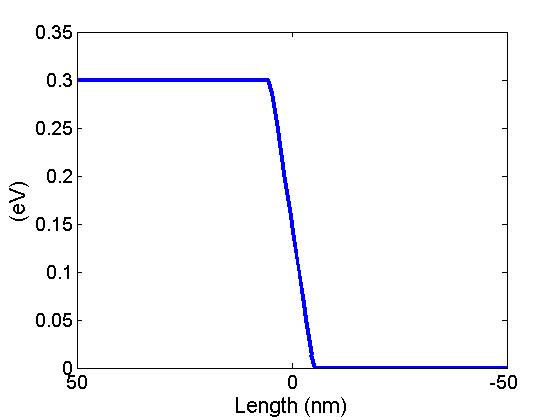}
b)\includegraphics[width=0.450\textwidth,height=0.37\textwidth]{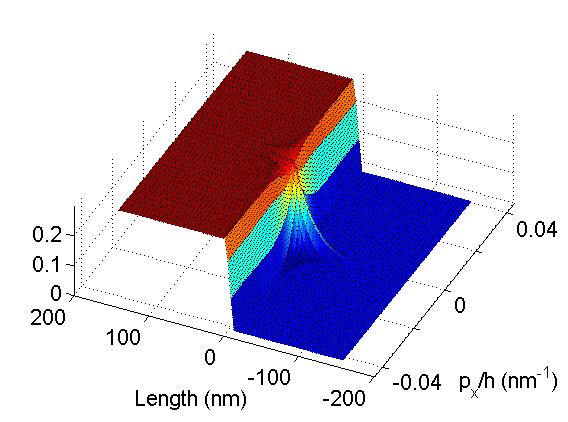}\\
c)\includegraphics[width=0.450\textwidth,height=0.37\textwidth]{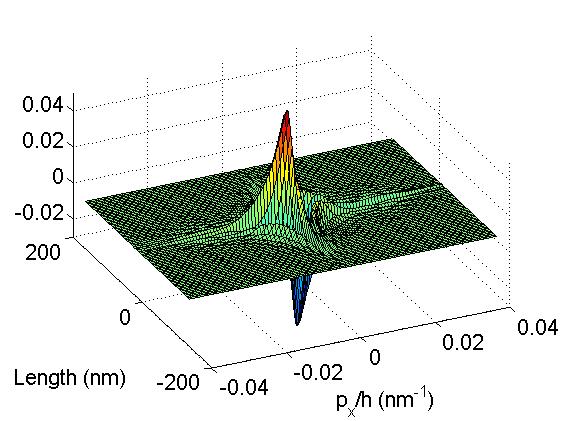}
d)\includegraphics[width=0.450\textwidth,height=0.37\textwidth]{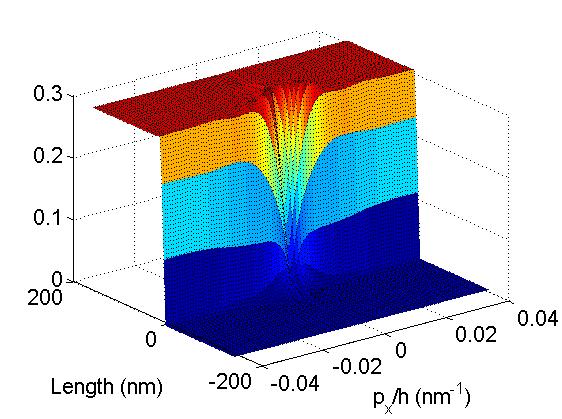}
\caption{Pseudo-potential:  a) External potential $U(\mathbf{r})$; b) $[\mathcal{U}'(\mathbf{r},\mathbf{p})]_{++}$; c) $[\mathcal{U}'(\mathbf{r},\mathbf{p})]_{+-}$; d) $[\mathcal{U}'(\mathbf{r},\mathbf{p})]_{--}$. Here $p_y/\hbar=10^{-3}$ nm$^{-1}$ .
}\label{fig pseudo_pot}
\end{center}
\end{figure}
\begin{figure}[!t]
\begin{center}
a)\includegraphics[width=0.450\textwidth,height=0.37\textwidth]{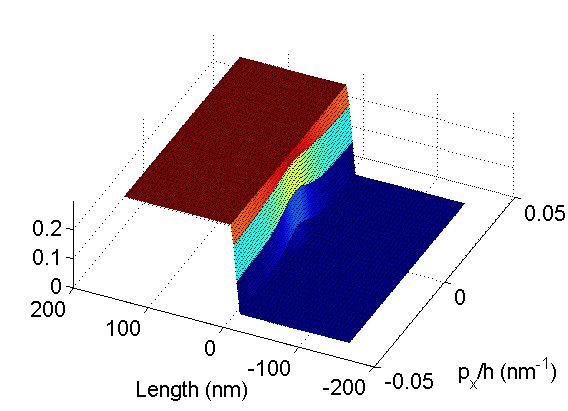}
b)\includegraphics[width=0.450\textwidth,height=0.37\textwidth]{pot_py_001_b.jpg}\\
c)\includegraphics[width=0.450\textwidth,height=0.37\textwidth]{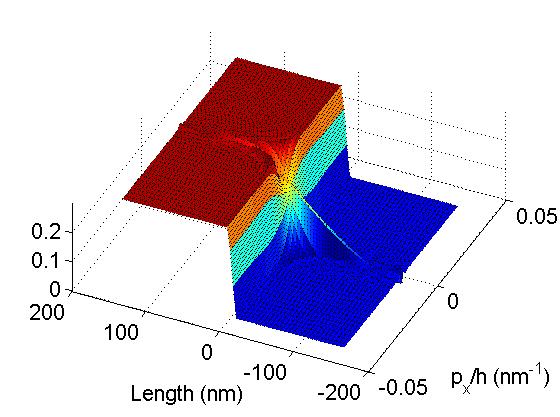}
d)\includegraphics[width=0.450\textwidth]{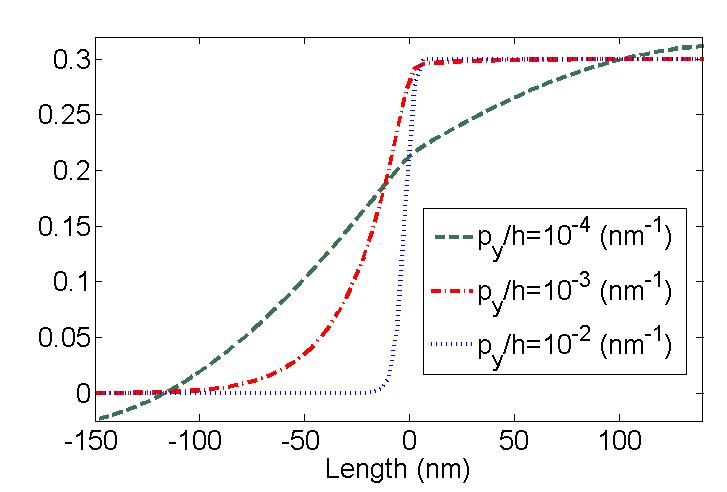}
\caption{$[\mathcal{U}'(\mathbf{r},\mathbf{p})]_{++}$ component of the pseudo-potential for different values of the momentum $p_y$:  a) $p_y/\hbar=10^{-2}\; \textrm{nm}^{-1}$; b) $p_y/\hbar=10^{-3}\; \textrm{nm}^{-1}$ c) $p_y/\hbar=10^{-4}\; \textrm{nm}^{-1}$. d) Pseudo-potential $[\mathcal{U}'(\mathbf{r},\mathbf{p})]_{++}$ for $p_x=0$.
}\label{fig pseudo_pot py var}
\end{center}
\end{figure}
In order to illustrate the main features of the pseudo-potential $\,\mathcal{U}'(\mathbf{r},\mathbf{p})$, in fig. \ref{fig pseudo_pot} we depict the explicit form of $\,\mathcal{U}'$ when the external potential $U(\mathbf{r})$ (represented in the sub-plot \ref{fig pseudo_pot}-a) is a barrier. Form Eq. \eqref{rela fra sym} we recognize that  $\mathcal{U}'$ is a $2\times 2$ matrix depending both on the position $\mathbf{r}$ and the momentum $\mathbf{p}$.
We note that some $\mathbf{p}$-dependent corrections to the potential arise around $p_x=0$, whereas the pseudo-potential stays practically identical to $U$ for greater values of the momentum. We remark that this characteristic reflects the presence of a singular behavior of the particle-hole motion in the proximity of the Dirac point. It will be addressed in more details in the following sections. In order to highlight the modification of the pseudo-potential when the parallel momentum $p_y$ changes, in fig. \ref{fig pseudo_pot py var}, we depict the  component $[\mathcal{U}'(\mathbf{r},\mathbf{p})]_{++}$ (in a single band description of the dynamics, it represents the potential ``seen" by the particles in the $\Sigma^+$ band) for different $p_y$. The plot shows that for large values of $p_y$, the in-band component of the pseudo-potential coincides with the external potential $U(\mathbf{r} )$. For small values of $p_y$, we note that the original step-like shape of the potential changes dramatically around $p_x=0$.  There, it becomes smoother and enlarges the spatial region where the gradient of the pseudo-potential (representing an effective electric field) differs from zero. This can be seen as a strong non-locality of the potential (or equivalently of the electric field) that is  a peculiarity of the graphene band structure and reflects the property that a particle around the Dirac point is quite immune to the localization effects. In our formalism, we describe this behavior by the presence of an effective potential $\mathcal{U}'$ that becomes more and more non-local when $|\mathbf{p}|$ goes to zero. This explains way, differing from the scattering process in normal metals, the transport of Dirac fermions in graphene sheets is only weakly affected by external sources of disorder (defects or impurities).
The behavior of the pseudo-potential $\mathcal{U}'$ around $|\mathbf{p}|=0$ can be investigated also analytically. By using that
\begin{eqnarray*}
\Theta ({\za^+}) \Theta^\dag ({\za^-})  &=&   \frac{1}{ 2   }  \left(
                                \begin{array}{cc}
                                 1+  e^{i(\zp_{\za^-}-\zp_{\za^+})}   & \displaystyle e^{-i\zp_{\za^+}} (e^{-(i\zp_{\za^-}-\zp_{\za^+})} - 1)   \\
                                 \displaystyle  e^{ i\zp_{\za^+}} (1 -  e^{ i(\zp_{\za^-}-\zp_{\za^+})}  )   &   1+  e^{-i(\zp_{\za^-}-\zp_{\za^+})}      \\
                                \end{array}
                              \right),
\end{eqnarray*}
where we applied the polar notation $\za^{\pm}=\zr_{\za^{\pm}}e^{i \zp_{\za^{\pm}}} \equiv p_x\pm \frac{\hbar}{2} \zm_x+ip_y$, it is easy to see that
\begin{eqnarray*}
\lim_{|\mathbf{p}|\rightarrow 0} \mathcal{U}'&=&  \left(
                                \begin{array}{cc}
                                 0   & - U(\mathbf{r})   \\
                                 U(\mathbf{r}) & 0  \\
                                \end{array}
                              \right),
\end{eqnarray*}
so that the in-band component of the pseudo potential vanishes. We remark that this consideration should not suggest that in the single-band limit the effect of the potential around the Dirac point vanishes and the particles move freely. Observing the equation of motion \eqref{[U S]} reveals that the pseudo-potential $\mathcal{U}'$ is non-local both in position and momentum, so that the particle motion is conditioned by the values of $\mathcal{U}'$ in an interval  of the momentum axes and not just at a point. In particular, only for spatially uniform Wigner distribution functions the pseudo-potential becomes local-in-momentum and its single band effect vanishes.

The principal aim of this contribution is to study the effect of the band-to-band transition on the stationary current in a graphene sheet. The full quantum mechanical description of motion consist of a rather complex set of coupled equations, where a simple interpretation of the dynamics is hampered by the presence of the highly non-local operators. In order to get more physical insight and to profit of the close analogy between the classical mechanics and the Wigner formalism, we consider the so-called gradient expansion procedure. We thus expand the functions $\Lambda$ and $\mathcal{U}'$ in Eqs. \eqref{[Lam S]}-\eqref{[U S]} with respect to $\hbar$ and  limit ourselves to the leading order. The study of the full quantum Wigner transport will be addressed in sec. \ref{sec num sol full quant}. If we expand the function $\Lambda$ up to the  first order in $\hbar$, Eq. \eqref{[Lam S]} simplifies to
\begin{eqnarray}
\left[\Lambda(\mathbf{p}) , \mathcal{S}' \right]_\star
&=&\left[\Lambda  ,\mathcal{S}' \right] - \frac{i\hbar }{ 2   }\left\{ \nabla_\mathbf{p} \Lambda , \nabla_\mathbf{r}  \mathcal{S}' \right\}
 + o( \hbar^2),   \label{[Lam S] pr ord}
\end{eqnarray}
where curly brackets denote the anti-commutator. In the hypothesis that the external electric potential $U(\mathbf{r})$ is regular, we have
\begin{eqnarray}
\left[\mathcal{U}' , \mathcal{S}' \right]_\star
&=&    i\hbar  \nabla_\mathbf{r}   {U} \cdot \nabla_\mathbf{p}  \mathcal{S}' + \frac{i\hbar}{2} \left[\left[ \Theta  , \nabla_{\mathbf{p}} \Theta  \cdot \nabla_{\mathbf{r}}       U       \right], \mathcal{S}'\right]   +o(\hbar^2)\label{[U S] pr ord} \; .
\end{eqnarray}
This approximation is justified in the limit where the external electric potential $U(\mathbf{r})$ can be considered as a sufficiently smooth function, so that only the first-order terms (proportional to the electric field) play a significant role in the dynamics.
In Eq. \eqref{[U S] pr ord} it is easy to identify the first term with the usual force operator. The second term takes the main quantum correction to the classical equation of motion into account. In the following, we will describe its physical meaning in terms of the band-to-band transition and exploit its connection with the adiabatic Berry phase approximation.
The equations of motion \eqref{mot Neum sec} become (the components of $\mathcal{S}'$ are defined in Eq. \eqref{comp Sp})
\begin{eqnarray}
\dpp{ f^\pm }{t}   &=& \pm \frac{v_F}{\sqrt{1+\xi^{-2}} } \frac{\mathbf{p}}{|\mathbf{p}|} \cdot \nabla_{\mathbf{r}} f^\pm+ \nabla_\mathbf{r}   {U} \cdot \nabla_\mathbf{p}  f^\pm \pm i \left(  \mathcal{B}    f^i -  \overline{ \mathcal{B} f^i} \right) \label{mot fpm}, \\
\dpp{ f^i }{t}   &=& i \mathcal{A}  f^i  +\nabla_\mathbf{r}   {U} \cdot \nabla_\mathbf{p}  f^i +i \overline{\mathcal{B}}  \left(  f^+ -  f^- \right), \label{mot fi}
\end{eqnarray}
where overbar means conjugation and
\begin{eqnarray}
  \mathcal{A} &=& -\frac{ 2 v_F }{ \hbar \sqrt{1+\xi^{-2}}} |\mathbf{p}| + \frac{(\mathfrak{M}^-)^2 }{ |\mathbf{p}|^2}   \left(\mathbf{p} \wedge \nabla_\mathbf{r} U   \right)_z  \; , \label{def coef A}\\
  \mathcal{B}   &=&  \frac{ \mathfrak{M}^+\mathfrak{M}^- }{2 }    \frac{p_x+ip_y}{|\mathbf{p}|^3}\left(\mathbf{p} \wedge \nabla_\mathbf{r} U   \right)_z  \; . \label{def coef B}
\end{eqnarray}
Here, $\mathfrak{M}^\pm (\xi)  = \sqrt{ 1 \pm \frac{1}{\sqrt{1+\xi^2} } }$, $\xi= \frac{v_F |\mathbf{p}|}{\Delta}$ and $\left(\mathbf{p} \wedge \nabla_\mathbf{r} U   \right)_z$ denotes the out-of-plane component ($z$-coordinate) of the vector $\left(\mathbf{p} \wedge \nabla_\mathbf{r} U   \right)$.%

Equations \eqref{mot fpm}-\eqref{mot fi} extend the semi-classical evolution of a two-particle system in a quantum mechanical context. In particular, in the limit of a vanishing electric field, the equations decouple and, as expected from a physical point of view, the particle system is described by two semi-classical equations of motion. This simple limit eases to attach a physical meaning to the various components of the solution. In particular, $f^+$ ($f^-$) represents the electron (hole) distribution function in the presence of an external electric field that modifies the crystal periodic potential (more precisely, they are the components of the Wigner function in a basis, where the two-band Hamiltonian is locally diagonal in the momentum and in the position space up to the first order in $\hbar$).

To appreciate the advantage of using our quasi-diagonal formalism, we compare the equations of motion \eqref{mot fpm}-\eqref{mot fi} with the evolution equations obtained by a direct application of the standard two-band Wigner formalism. The Wigner function for a multiband system is usually defined as 
\begin{eqnarray}
\mathfrak{f}_{ij}(\mathbf{r},\mathbf{p})
=    \frac{1}{(2\pi\hbar)^2}  \int  \psi_i \left(\mathbf{r}+\frac{\bs{\zh}}{2}\right) \psi_j  \left(\mathbf{r}-\frac{\bs{\zh}}{2}\right) e^{-\frac{i}{ {\hbar}} \mathbf{p} \cdot \bs{\zh}}\dif \bs{\zh} \; , \label{wigner transf}
\end{eqnarray}
where $\Psi = { \psi_1  \choose  \psi_2 }$ is the two component Schr\"odinger wave function satisfying $ i \hbar \dpp{\Psi}{t} = \mathcal{H} \Psi$. Equation \eqref{wigner transf} is a straightforward extension of the single band Wigner function, where the Wigner transformation is applied componentwise to the density matrix. Up to the first order in $\hbar$, the equation of motion for the two-component Wigner function writes
 \begin{eqnarray}
 \dpp{\mathbf{f}^S\left(\mathbf{r} ,\mathbf{p} \right)}{t}   &=&  - \frac{v_F   }{ 2    }  \nabla_{\mathbf{r}} {f}_0  + \left(\nabla_\mathbf{r}   {U} \cdot \nabla_\mathbf{p} \right) \mathbf{f}^{S}  + \frac{v_F}{\hbar} \mathbf{f}^S \wedge \mathbf{p} \; , \label{spin wig vf}   \\
 \dpp{ f_0 \left(\mathbf{r} ,\mathbf{p} \right)}{t}   &=&   \nabla_\mathbf{r}   {U} \cdot \nabla_\mathbf{p}  f_0 - \frac{ v_F }{ 2    }  \; \mathrm{div}\mathbf{f}^S   \; , \label{spin wig f0}
\end{eqnarray}
where we defined the vector $\mathbf{f}^S =   \left(2 \Re \left\{ \mathfrak{f}_{21}\right\} , 2 \Im \left\{ \mathfrak{f}_{21}\right\} , \mathfrak{f}_{11}-\mathfrak{f}_{22} \right) $, $f_0 = \mathfrak{f}_{11}+ \mathfrak{f}_{22}$ and $\Im$ ($\Re$) denotes the imaginary (real) part.
The formulation of the two-band Wigner approach given in Eqs. \eqref{spin wig vf}-\eqref{spin wig f0} is characterized by the presence of high oscillating regimes. The direct numerical treatment of Eqs. \eqref{spin wig vf}-\eqref{spin wig f0} reveals itself to be a very difficult task. The quantum mechanical two-band motion is essentially a two-scale process characterized by band-to-band transitions (whose frequency is proportional to the energy difference between states localized in the upper and the lower Dirac cones) and the intraband motion of the electrons (that, with respect to the tunneling processes, can be considered as a slow dynamical process).
Furthermore, in this formulation the analogy with the semi-classical evolution of the system (characterized by two uncoupled Liouville equations, one for the particle distribution function in the upper cone, and one for the hole distribution function in the lower cone) is completely lost. Here, a description of the dynamics where we can associate a certain quasi-distribution function to the particle and a different quasi-distribution function to the holes, does not apply.
One of the most remarkable advantages of the single-band Wigner formulation of the quantum mechanics (and was the main reason for which this formulation has been introduced) is that in this framework the classical limit $\hbar\rightarrow 0 $ is easily evaluated. As shown by the Eqs. \eqref{spin wig vf}-\eqref{spin wig f0}, this is no longer true in the many-band case, where the limit $\hbar\rightarrow 0 $ is completely non-trivial. This is due to the presence of the last term of Eq. \eqref{spin wig vf}. When $\hbar$ goes to zero, the various components of $f$ become more and more coupled and the system becomes ill defined. This simple consideration suggests to use instead of $(\mathbf{f}^S,f^0)$, some new unknowns behaving regularly in the limit $\hbar\rightarrow 0$. This can be obtained by the partial diagonalization procedure described in this section.

\subsection{Berry connection in the quantum phase space}\label{sec berry phase}

In a crystal where the effective Hamiltonian is expressed by a partially diagonalized basis (such as for example graphene or Kane-Luttinger $kp$ models for semiconductors), the velocity operator has off-diagonal elements and the electric field mixes the bands, so that the expectation value of the velocity acquires an additional term proportional to the field and the usual definition of group velocity does no longer apply. The theory of Berry phases offers an elegant explanation of this effect in terms of the intrinsic curvature of the perturbed band. Furthermore, the Berry connection plays an important role in spin dynamics and in describing spin-orbit interactions. We discuss how it is possible to characterize the Berry phase in graphene (which is usually studied at the Schr\"odinger level) by using our kinetic description of the quantum dynamics. The formal analogy between spin and band degree of freedom suggests that we investigate the effects of including the Berry phase in the evolution of a many-band electron system. We apply the Berry approach to the Hamiltonian symbol $\mathcal{H}$ (which is a simple matrix where $\mathbf{p}$ plays the role of the adiabatic variable). Berry's adiabatic theory states that, if a system is initially described by a certain eigenvector $u_i (\mathbf{p})$ of $\mathcal{H}(\mathbf{p})$, the vector state of the system at time $t$ is given by
\begin{eqnarray}
\psi (t) =u_i (\mathbf{p}(t))  \;  e^{i \gamma_i (t)-\frac{i}{\hbar} \int_0^t \ze_i[\mathbf{p}(t')] \dif t' } \; ,  \label{Berry sol}
\end{eqnarray}
where the term $\gamma_i$ is named dynamical phase factor and can be obtained as the path integral along the $\mathbf{p}$-trajectory,
$
\gamma_i  = \int \mathbf{A}_{ii}(\mathbf{p}) \cdot \dif \mathbf{p}
$, 
of the Berry connection $\mathbf{A}(\mathbf{p})$  given by
$
  \mathbf{A}_{ij}(\mathbf{p}) = i \left\langle u_i (\mathbf{p}) | \nabla_{\mathbf{p}}  u_j (\mathbf{p}) \right\rangle
$. 
In our case, by construction, the distribution functions $f^+$ and $f^-$, respectively, are the Wigner functions related to the $\mathbf{p}$-dependent Floquet projectors $\left| u_+ (\mathbf{p}) \right\rangle \left\langle \; u_+ (\mathbf{p}) \right| $ and $\left| u_-(\mathbf{p})  \right\rangle \left\langle \; u_-(\mathbf{p})  \right| $. Since from Eq. \eqref{Berry sol} we have that
\begin{eqnarray*}
\left| \psi (t) \right\rangle \left\langle  \psi (t) \right| = \left| u_i [\mathbf{p} (t)] \; \right\rangle \left\langle \; u_i [\mathbf{p}(t)] \; \right|
\end{eqnarray*}
for these functions, the Berry phases cancel out. On the contrary, the function $f^i$ is related to the ``band transition" operator $\left| u_+(\mathbf{p})  \right\rangle \left\langle \; u_-(\mathbf{p})  \right| $ that, for a given trajectory $\textbf{p}(t)$, cumulates a Berry phase equal to
\begin{eqnarray}
&& \dpt{\mathbf{p}}{t} \cdot \left( \mathbf{A}_{++}-\mathbf{A}_{--} \right)  -\frac{ \ze_+(\mathbf{p})-\ze_-(\mathbf{p} ) }{\hbar}
=  \frac{(\mathfrak{M}^-)^2 }{ |\mathbf{p}|^2}   \left(\mathbf{p} \wedge  \dpt{\mathbf{p}}{t}    \right)_z    -\frac{ 2 v_F }{ \hbar \sqrt{1+\xi^2}}  |\mathbf{p}|=\mathcal{A} , \label{Berry non diag}
\end{eqnarray}
where $\dpt{\mathbf{p}}{t} =\nabla_{\mathbf{r}} U  $. We see that the Berry phase coincides with the ``natural" oscillation frequency of $f^i$. Our method is thus particularly suited to highlight the  role of the Berry phase in the evolution of the system. A well known characteristics of the Berry connection is the divergence in the proximity of points where the bands intersect. In gapless graphene, such a divergence can be found in the neighborhood of the Dirac point $\mathbf{p}=0$. For that reason, from Eq. \eqref{def coef A} we see that the natural oscillation frequency $\mathcal{A}$ of $f^i$ behaves like $1/|\mathbf{p}|$ when $\Delta=0$.

\section{Simulation of graphene (uniform electric field)}\label{sec unif ele}
The most common configuration to perform experiments with graphene is constituted by the single-layer graphene field effect transistor (FET) \cite{Zhang_09}. Graphene FETs are fabricated by standard lithography and a degenerately doped silicon substrate is used to tune the 2D carrier density in the proximity of the Dirac point.
We apply our model in the approximation of a quasi-uniform electric field (constituted by Eqs. \eqref{mot fpm}-\eqref{mot fi}) in order to study the quantum corrections to the ballistic charge motion in an intrinsic graphene sheet (for which $\Delta=0$ in Eqs. \eqref{mot fpm}-\eqref{mot fi}) induced by an applied external potential.
We prescribe boundary conditions  in correspondence to the metallic contacts. The contacts are considered as perfect charge reservoirs, where the number of particles entering the device are given by the thermal equilibrium distribution. By identifying $f^+$ and $f^-$ with the electron distribution functions in the upper ($\Sigma^+$) and lower ($\Sigma^-$) cone, respectively, we fix their incoming values at the boundaries of the simulation domain equal to the Fermi distribution function. Vanishing boundary conditions are assigned to the interband function $f^i$.

We consider a simple device consisting of a graphene sheet suspended by two ohmic contacts at a distance of $1\; \mu$m. The bias voltage $U$ is applied between the contacts. This prototype of devices has been experimentally probed in \cite{Bolotin_08}. The presence of interfacial phonons in the substrate reveals itself to be an important source of limitation for the charge mobility in graphene. However, suspended graphene offers the considerable advantage that the interactions between the underlying substrate and the graphene sheet are completely eliminated.
Up-to-date lithographic technique allows the fabrication of high quality graphene sheets suspended on a silicon substrate where the mean distance between the flat graphene sheet and the substrate is around 150 $nm$. Under this conditions, we can safely assume that no phonons are transmitted to the graphene sheet from the substrate. At room temperature, mobilities of suspended graphene are close to $10^4$ $\textrm{cm}^2 \textrm{V}^{-1} \textrm{s}^{-1}$, and are limited by acoustic phonon scattering. Mobilities of such an order of magnitude imply that electrons can travel from one contact to the other by suffering only a few scattering events. This evidence justifies the study of ballistic transport in suspended graphene. As a further simplification, we assume that the particles move under the action of an external electric field $\mathcal{E}$ directed along the $x$ direction and independent of the $y$ variable.
\begin{figure}[!t]
\begin{center}
 {\begin{minipage}[h]{0.48\textwidth}
\includegraphics[width=\textwidth,height=0.74\textwidth]{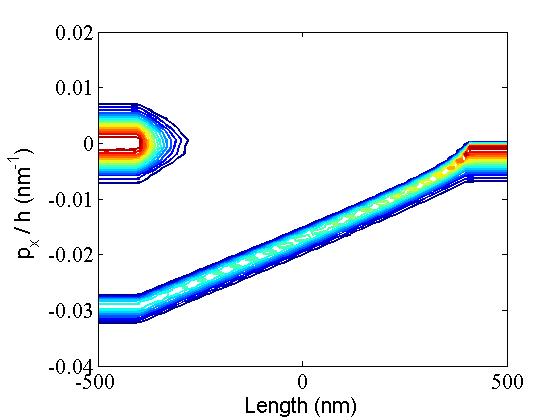} \\
\flushleft  \vspace{-12mm} a)
\end{minipage}}
 {\begin{minipage}[h]{0.48\textwidth}
\includegraphics[width=\textwidth,height=0.74\textwidth]{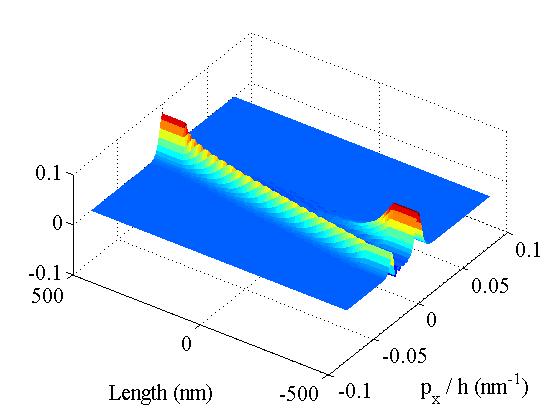} \\
 \flushleft \vspace{-12mm} b)
\end{minipage}}\\
 {\begin{minipage}[h]{0.48\textwidth}
\includegraphics[width=\textwidth,height=0.74\textwidth]{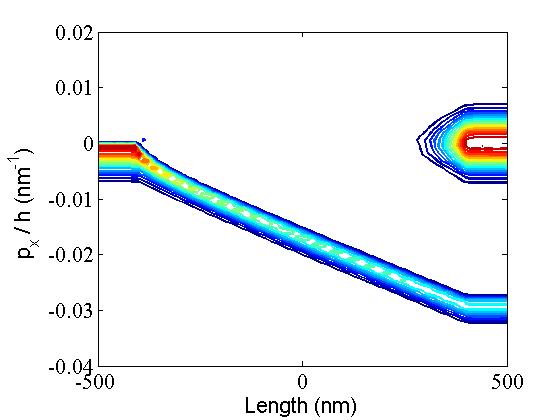} \\
\flushleft  \vspace{-12mm} c)
\end{minipage}}
 {\begin{minipage}[h]{0.48\textwidth}
\includegraphics[width=\textwidth,height=0.74\textwidth]{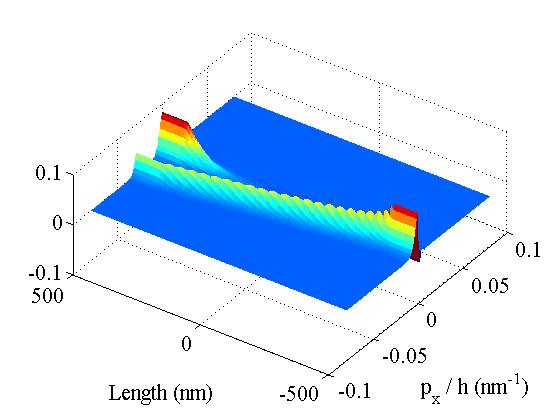} \\
 \flushleft \vspace{-12mm} d)
\end{minipage}}
\caption{Stationary solution for graphene for an applied potential $V_0=0.3$ eV. Snapshots of the $f^+$ (upper plot), $1-f^-$ (lower plot) distributions on the $x-p_x$ plane, for $p_y/\hbar = 0.1$ nm$^{-1}$. In the left plot we represent the contour lines and in the right plot the 3D representation of the solutions.}\label{fig grap fpm 2a}
\end{center}
\end{figure}
\begin{figure}[!t]
\begin{center}
 {\begin{minipage}[h]{0.48\textwidth}
\includegraphics[width=\textwidth,height=0.74\textwidth]{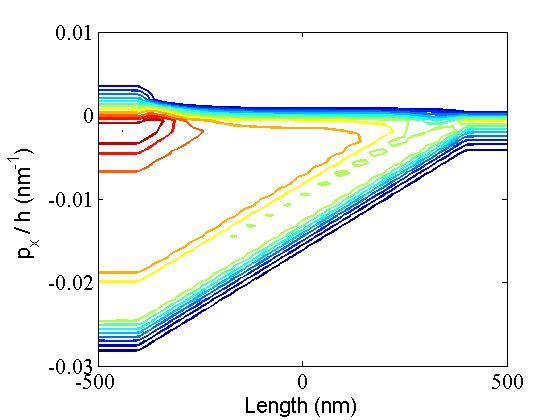} \\
\flushleft  \vspace{-12mm} a)
\end{minipage}}
 {\begin{minipage}[h]{0.48\textwidth}
\includegraphics[width=\textwidth,height=0.74\textwidth]{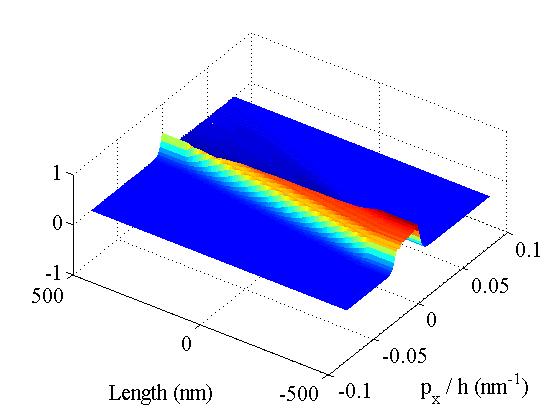} \\
 \flushleft \vspace{-12mm} b)
\end{minipage}}\\
 {\begin{minipage}[h]{0.48\textwidth}
\includegraphics[width=\textwidth,height=0.74\textwidth]{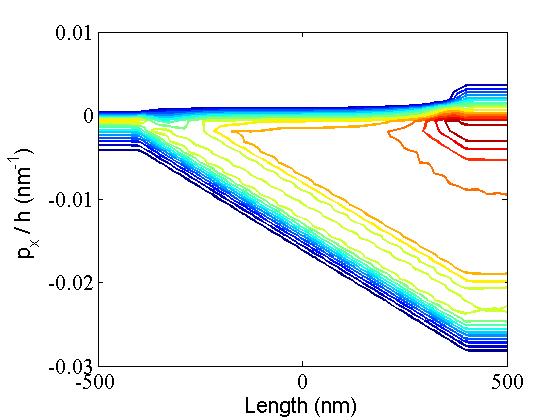} \\
\flushleft  \vspace{-12mm} c)
\end{minipage}}
 {\begin{minipage}[h]{0.48\textwidth}
\includegraphics[width=\textwidth,height=0.74\textwidth]{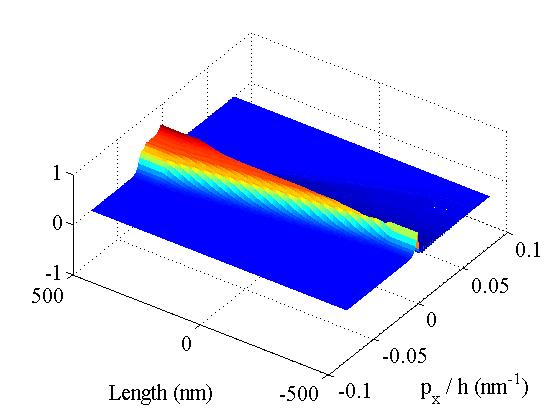} \\
 \flushleft \vspace{-12mm} d)
\end{minipage}}\\
 {\begin{minipage}[h]{0.48\textwidth}
\includegraphics[width=\textwidth,height=0.74\textwidth]{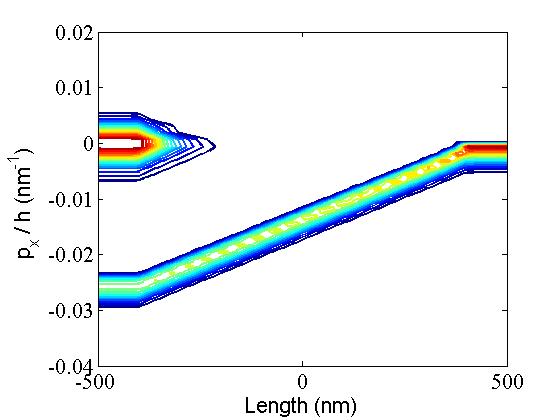} \\
\flushleft  \vspace{-12mm} e)
\end{minipage}}
 {\begin{minipage}[h]{0.48\textwidth}
\includegraphics[width=\textwidth,height=0.74\textwidth]{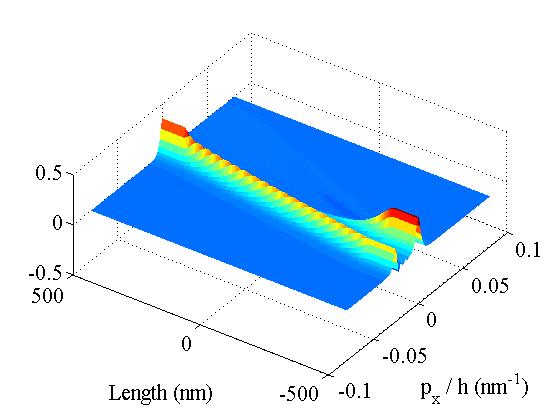} \\
 \flushleft \vspace{-12mm} f)
\end{minipage}}
\caption{Stationary solution for graphene under the external potential  $V_0=0.3$ eV. Snapshots of the $f^+$ (a-b) and $1-f^-$ (c-d) distributions on the $x-p_x$ plane for $p_y/\hbar= 2\cdot 10^{-3}\;\textrm{nm}^{-1}$. In the left plots we represent the contour lines and in the right plots the 3D representation of the solutions. In (e-f) we plot the semiclassical $f^+$ solution (without band transition).} \label{fig grap fpm 2} 
\end{center}
\end{figure}
In this case, we can assume $\dpp{f}{y}=0$ which greatly reduces the numerical complexity of the system. The equations of motion \eqref{mot fpm}-\eqref{mot fi} simplify to
\begin{eqnarray}
\dpp{ f^\pm }{t}   &=& \pm v_F \frac{p_x}{ \sqrt{ p_x^2+p_y^2} }   \dpp{ f^\pm}{x}+ \mathcal{E}  \dpp{f^\pm}{p_x} \pm \mathcal{E} \frac{p_y}{ p_x^2+p_y^2} \Im \left\{  \frac{p_x+ip_y}{\sqrt{p_x^2+p_y^2}} \; f^i\right\} \; ,  \nn \\ \label{fpm unidim} \\
\dpp{ f^i }{t}   &=& i \mathcal{A} f^i +\mathcal{E} \dpp{f^i}{p_x}    -i \frac{ \mathcal{E} }{2  }   \frac{p_y\left( p_x-ip_y\right)}{\left( p_x^2+p_y^2\right)^{3/2}}  \left(  f^+ -  f^- \right)  \; , \nn \\ \label{fi unidim}
\end{eqnarray}
where
\begin{eqnarray}
\mathcal{A}(p_x,p_y) &=&     - \mathcal{E}  \frac{p_y}{ p_x^2+p_y^2} -  \frac{2 v_F }{\hbar}  \sqrt{ p_x^2+p_y^2} \; ,  \label{def Q}
\end{eqnarray}
and $\mathbf{p}=(p_x,p_y)$, $\mathbf{r}=(x,y)$.
In the figs. \ref{fig grap fpm 2a}-\ref{fig grap fpm 2} we depict the stationary values of the electron  distribution $f^+$ and the hole distribution $(1-f^-)$ for an external applied potential $V_{0}=0.3$ eV for different values of the parallel momentum $p_y$. In the left plots we represent the contour lines and in the right plots the 3D representation of the solutions. In our simulation, the lattice temperature $T = 300 $ K. It can be clearly seen that the electrons split in two parts: the particles with velocities parallel to the electric field which are accelerated, and anti-parallel ones which are reflected back by
the potential barrier. Further, due to the presence of interband Klein tunneling, also interband particle transitions between the bands $\zS^+$ and $\zS^-$ are possible. Since the relation between the velocity and the momentum for a hole is the inverse of that for an electron during this interband transition, the momentum parallel to the barrier is conserved and the velocity of the quasiparticle is inverted. Due to the larger number of particles in the lower cone, we observe a net flux of particles from $\zS^-$ towards $\zS^+$.
As expected, interband transitions become a dominant phenomenon around the Dirac point $\mathbf{p}=0$. In fact, in correspondence to high values of $p_y$ (depicted in figure \ref{fig grap fpm 2a}), the distribution functions look very similar to their classical counterparts and quantum corrections are negligible. On the contrary, for smaller values of $p_y$  (see fig. \ref{fig grap fpm 2}) a flux of particles from the $\zS^-$ band to the $\zS^+$ band is clearly visible. To highlight the effect of Klein tunneling, in fig.  \ref{fig grap fpm 2}-e-f we present the distribution functions of electrons and holes under the same condition as in fig.  \ref{fig grap fpm 2}-a-b but in the semi-classical approximation (without tunneling). One of the advantages of our approach is that now the Klein tunneling effect can be described by the familiar language of classical mechanics. In fact, from fig.  \ref{fig grap fpm 2}-c-d we see that, in order to overcome the potential barrier applied between the two contacts, a large number of particles belonging to the $\Sigma^-$ band leave this band and a corresponding increase of the related hole distribution function ($1-f^-$) is observed. These particles are now accelerated by the same electric field in the final part of the device ($x=L$) and contribute to increase the particle distribution $f^+$.

\subsection{Current and density}\label{sec. mac quant}

\begin{figure}[!t]
\begin{center}
 {\begin{minipage}[h]{0.48\textwidth}
\includegraphics[width=\textwidth,height=0.74\textwidth]{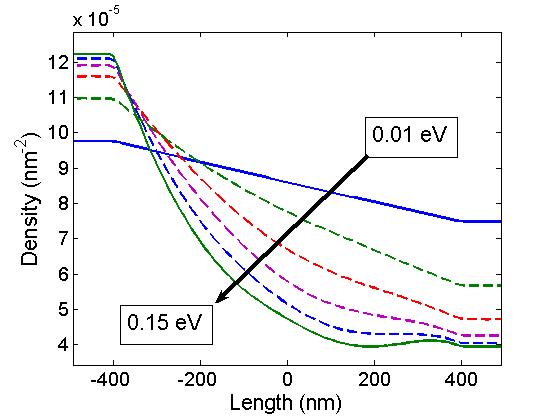} \\
\flushleft  \vspace{-12mm} a)
\end{minipage}}
 {\begin{minipage}[h]{0.48\textwidth}
\includegraphics[width=\textwidth,height=0.74\textwidth]{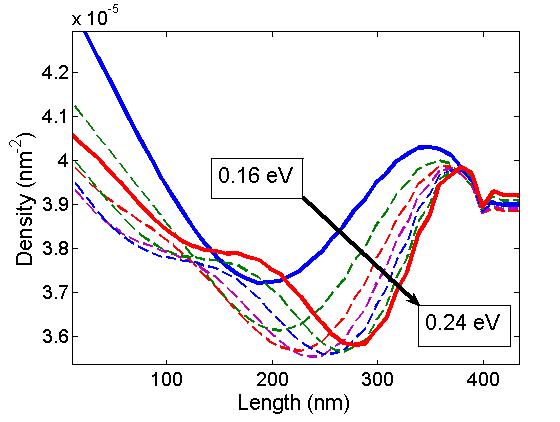} \\
 \flushleft \vspace{-12mm} b)
\end{minipage}}\\
 {\begin{minipage}[h]{0.48\textwidth}
\includegraphics[width=\textwidth,height=0.74\textwidth]{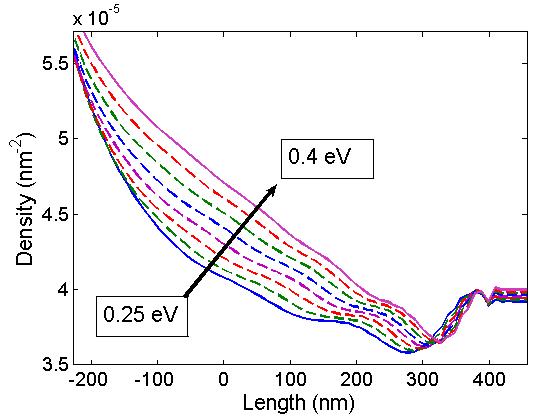} \\
\flushleft  \vspace{-12mm} c)
\end{minipage}}
 {\begin{minipage}[h]{0.48\textwidth}
\includegraphics[width=\textwidth,height=0.74\textwidth]{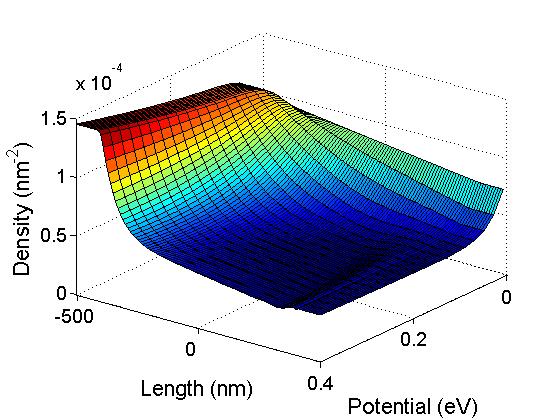} \\
 \flushleft \vspace{-12mm} d)
\end{minipage}}
\caption{a-c): Particle density in the $\Sigma^+$ band ($\mathfrak{n}^+$) for different applied potentials. d): 3D plot of $\mathfrak{n}^+$ as a function of the applied potential and the position. 
}
\label{fig dens tot}
\end{center}
\end{figure}

The density and current of particles in the upper (lower) band, denoted by $\mathfrak{n}^+$ ($\mathfrak{n}^-$) and $\mathfrak{j}^+$ ($\mathfrak{j}^-$), respectively, can be obtained from the Wigner functions as
\begin{eqnarray}
\mathfrak{n}^\pm(\mathbf{r},t) &=& \frac{1}{(2\pi\hbar )^2} \int f^\pm   (\mathbf{r},\mathbf{p},t) \dif \mathbf{p}\; , \label{npm density} \\
\mathfrak{j}^\pm(\mathbf{r},t) &=& \pm  \frac{e v_F}{(2\pi\hbar )^2} \int \frac{\mathbf{p}}{|\mathbf{p}|} f^\pm   (\mathbf{r},\mathbf{p},t) \dif \mathbf{p} \;   . \label{Jpm current}
\end{eqnarray}
The continuity equation for the charge can be deduced from the system of Eqs. (\ref{mot fpm})-(\ref{mot fi}):
\begin{eqnarray}
\dpp{ \mathfrak{n}^\pm }{t}   &=&   \nabla_{\mathbf{r}}\cdot  \mathfrak{j}^\pm   \mp \int\mathcal{M}[f^i] \dif \mathbf{p}  \; ,
 \label{continuity eq}\\
\mathcal{M}[f^i]  &=& \frac{1 }{ p^2} \left(\mathbf{p} \wedge \nabla_\mathbf{r} U   \right)_z  \Im \left\{f^i e^{  i\theta_{\mathbf{p}}} \right\}\; .
\end{eqnarray}
In particular, corresponding to a stationary solution (where $\dpp{ \mathfrak{n}^\pm }{t} =0 $), the total current $\mathfrak{j}^t \equiv\mathfrak{j}^+ + \mathfrak{j}^- $ becomes uniform ($ \nabla_{\mathbf{r}}  \mathfrak{j}^t =0 $).
Figure \ref{fig dens tot} shows the stationary charge density profile $\mathfrak{n}^+(x)$  in the intrinsic graphene for different applied voltages.
For low voltages, the behavior of the particle density is essentially semi-classical: with the increase of the external field, the electrons cumulate near the source contact and a charge depletion in the channel is observed. The nearly total depletion of the drain contact is reached for an applied potential of $0.15$ eV (snapshot $a$ of fig. \ref{fig dens tot}). In correspondence to a further increase of the applied potential, the  quantum Klein effect starts to play a relevant role in the shape settlement of the stationary density profile. In particular, for a potential greater than $0.25$ eV (snapshot $c$ of figure \ref{fig dens tot}), we observe a monotone increase of the charge density inside the channel.
This effect is due to the particles, initially localized in the $\Sigma^-$ band, that are injected in the upper cone $\Sigma^+$ as a response to such a strong electric field. We see that in this regime of a strong external potential, the quantum correction to the density becomes comparable with the total charge present in the device. Finally, in fig. \ref{fig dens tot}-$b$ we highlight the presence of density oscillations in the proximity of the ohmic contact.

\begin{figure}[!h]
\begin{center}
\includegraphics[width=0.450\textwidth,height=0.37\textwidth]{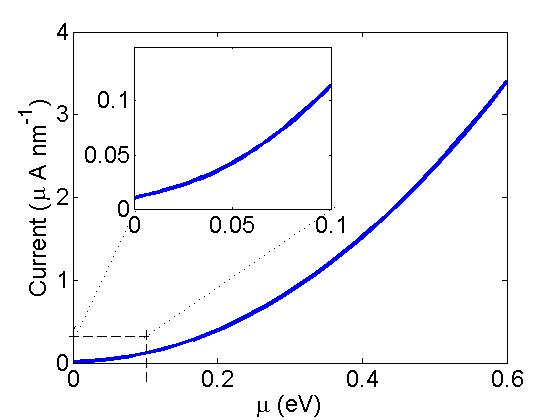}%
\caption{Saturation current $J_{sat}$ versus the chemical potential $\mu$ for a temperature of $300$ K. $J_{sat}$ is evaluated by Eq. (\ref{sat curr}). 
}\label{fig sat cur vs mu}
\end{center}
\end{figure}

We focus now our attention on the quantum correction to the total stationary current. It is well know that, when scattering processes are neglected, no steady state can exist in the graphene bulk. If a uniform electric field is applied to a spatially infinite sheet,
the momentum of the particles would increase indefinitely and the current would show Bloch oscillations \cite{Rosenstein_10}.
Nevertheless, when the real band structure of the graphene is approximated by an unbounded bi-conical shape, the saturation of ballistic current is reached. In fact, no matter how much they are accelerated, particles produce always the same amount of current. The upper limit of the current is obtained when all the particles entering the device through the source contact reach the drain and, at the same time, the drain incoming particles are reflected by the barrier. The current ($J_{sat}$) and density ($n_{in}$) related to the incoming particles distribution $f_{in}$ at the source contact are
\begin{eqnarray}
  n_{in}   &=& \frac{1}{(2\pi\hbar )^2} \int |\mathbf{p}|  f_{in}  (|\mathbf{p}|,\zf) \dif \zf \dif |\mathbf{p}|=  \frac{1}{4\pi}   \left(\frac{m_0 v_F}{ \hbar }\right)^2  \int \zr \left[1+ e^{\zb\left( \zr m_0 v_F^2 - \mu \right)  } \right]^{-1}  \textrm{d} \zr \;, \\
J_{sat} &=&  - \frac{ 2 e_0 v_F}{ \pi}  n_{in}.  \label{sat curr}
\end{eqnarray}
In figure \ref{fig sat cur vs mu} we plot the saturation current $J_{sat}$ versus the chemical potential $\mu$ for a temperature of $300$ K. 
The ballistic saturation current shows a considerable increasing when the chemical potential is augmented. On the contrary, numerical simulations proof that the quantum correction to the total current induced by interband tunneling is almost insensitive to a variation of the chemical potential. This can be understood if we note that the Klein tunneling in presence of a (almost uniform) slowly varying electric field concerns particles whose energy is located around the Dirac point (or equivalently, particles whose momentum is nearly zero). For low temperature, if the chemical potential is above the Dirac point, the number of such particles is almost independent from $\mu$. In this contribution, we will focus our attention to a quasi-intrinsic graphene sheet, for which the interband current is of the same order of magnitude as the saturation current.
\begin{figure}[!t]
\begin{center}
\includegraphics[width=0.450\textwidth,height=0.37\textwidth]{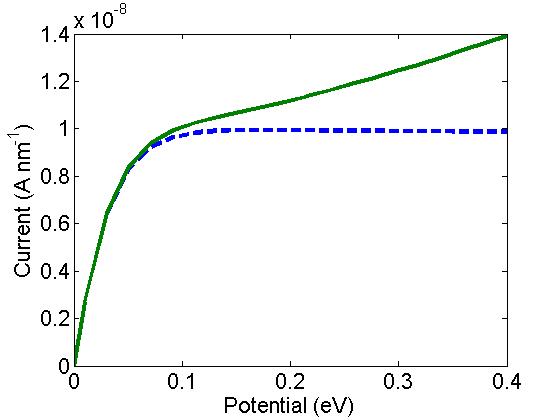}
\caption{I-V Characteristic: Comparison of the quantum solution (continuous green line) with the semi-classical solution (dashed blue line).
}\label{fig curr tot}
\end{center}
\end{figure}

The stationary $I-V$ characteristic of the device (intrinsic graphene) at the temperature of $300$ K is depicted in fig. \ref{fig curr tot}, where the current $I$ flowing through the device is plotted as a function of the bias voltage $V_{0}$ applied between the source and the drain contacts. We compare the solution of our quantum system with the classical motion. For this purpose, we plot the results obtained by discarding the interband transitions (dashed blue line) and including the multiband corrections (continuous green line). Our simulations show the importance of including the tunneling process in a realistic simulation of the current in intrinsic graphene. In particular, in the case of quantum transport, no saturation is observed and the current grows with increasing external potential.

\subsection{Non intrinsic graphene}
\begin{figure}[!h]
\begin{center}
 {\begin{minipage}[h]{0.48\textwidth}
\includegraphics[width=\textwidth,height=0.74\textwidth]{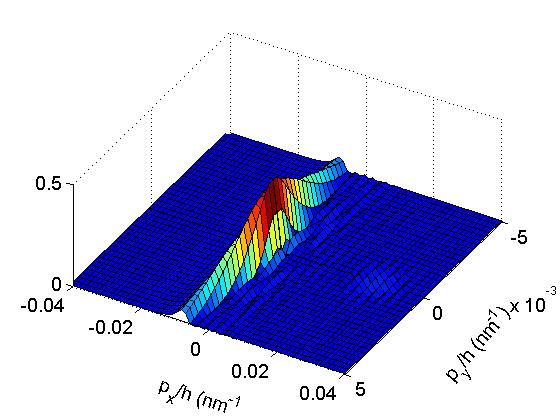} \\
\flushleft  \vspace{-12mm} a)
\end{minipage}}
 {\begin{minipage}[h]{0.48\textwidth}
\includegraphics[width=\textwidth,height=0.74\textwidth]{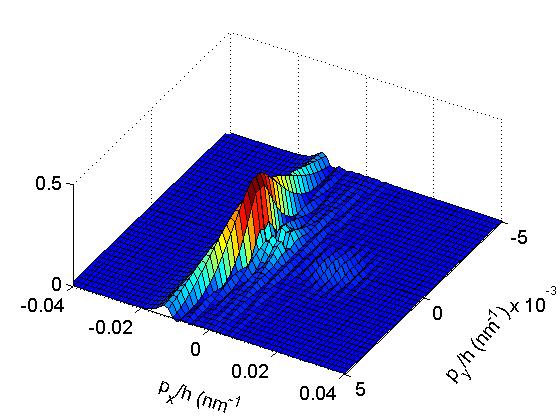} \\
 \flushleft \vspace{-12mm} b)
\end{minipage}}\\
 {\begin{minipage}[h]{0.48\textwidth}
\includegraphics[width=\textwidth,height=0.74\textwidth]{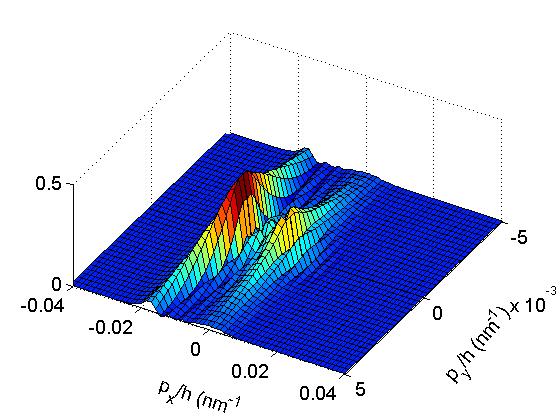} \\
\flushleft  \vspace{-12mm} c)
\end{minipage}}
 {\begin{minipage}[h]{0.48\textwidth}
\includegraphics[width=\textwidth,height=0.74\textwidth]{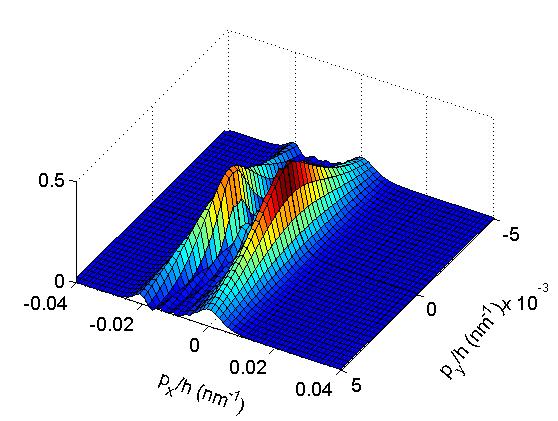} \\
 \flushleft \vspace{-12mm} d)
\end{minipage}}
\caption{Stationary solution for an applied potential of $V_0=0.3$ eV and $\mu=0$. Snapshot of the $f^+$ distributions in the plane $p_x-p_y$, at different positions $x$ along the device: (a) $x=L$ (source contact)   (b) $x=L/3$ (c) $x=-L/3$ (d) $x=-L$ (drain contact).
}\label{fig grap fpm pxy 1}
\end{center}
\end{figure}
\begin{figure}[!h]
\begin{center}
 {\begin{minipage}[h]{0.48\textwidth}
\includegraphics[width=\textwidth,height=0.74\textwidth]{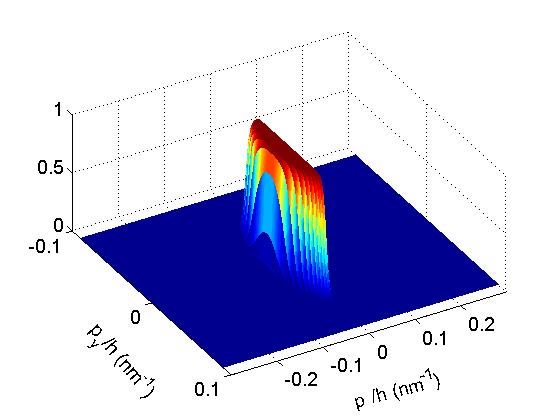} \\
\flushleft  \vspace{-12mm} a)
\end{minipage}}
 {\begin{minipage}[h]{0.48\textwidth}
\includegraphics[width=\textwidth,height=0.74\textwidth]{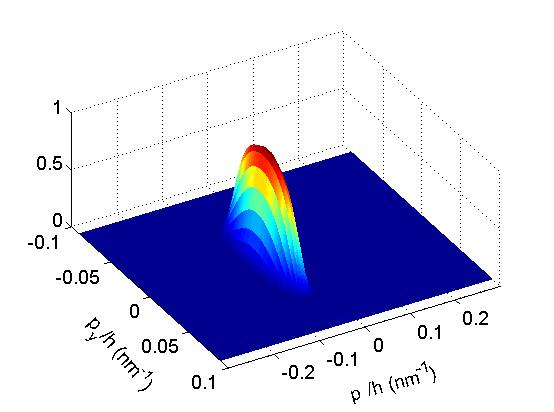} \\
 \flushleft \vspace{-12mm} b)
\end{minipage}}\\
 {\begin{minipage}[h]{0.48\textwidth}
\includegraphics[width=\textwidth,height=0.74\textwidth]{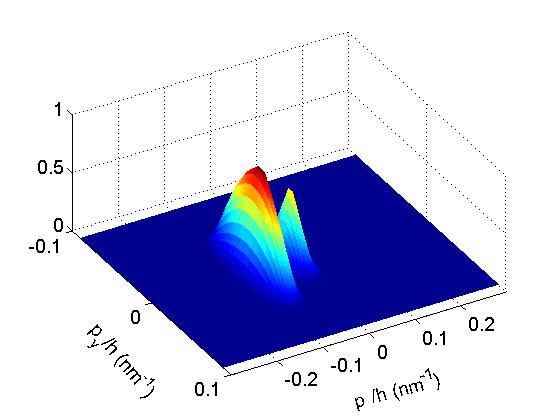} \\
\flushleft  \vspace{-12mm} c)
\end{minipage}}
 {\begin{minipage}[h]{0.48\textwidth}
\includegraphics[width=\textwidth,height=0.74\textwidth]{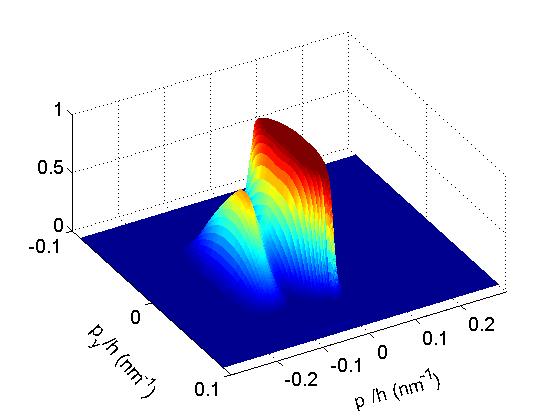} \\
 \flushleft \vspace{-12mm} d)
\end{minipage}}
\caption{Stationary solution for graphene under the external potential  $V_0=0.3$ eV and $\mu=0.6$ eV. The snapshots represent $f^+$ for different positions $x$ along the device : (a) $x=L$ (source contact)   (b) $x=L/3$ (c) $x=-L/3$ (d) $x=-L$ (drain contact). }\label{fig grap fpm 3}
\end{center}
\end{figure}
In order to give a clearer description of the two-band motion and to compare the solution of intrinsic graphene with doped graphene, we represent the solutions $f^\pm$ in the $p_x-p_y$ plane. In particular, in fig. \ref{fig grap fpm pxy 1} we represent the stationary solution for the distribution function $f^+$ in intrinsic graphene under the same condition as in the figs. \ref{fig grap fpm 2a}-\ref{fig grap fpm 2} (the external potential $V_0=0.3$ eV). The snapshots show the $f^+$ distribution at different positions $x$ along the device: (a) $x=L$ (source contact), (b) $x=L/3$, (c) $x=-L/3$, (d) $x=-L$ (drain contact).
We see that the $\Sigma^+$ particles entering the device from the source contact are accelerated by the potential and leave the device at $x=-L$ without reflection. On the contrary, particles injected in the graphene sheet from the drain contact have not enough energy to overcome the potential barrier and are reflected. Based on these general considerations, we see that in the snapshot of fig. \ref{fig grap fpm pxy 1}-a) only one electron beam is visible. The following cuts of the solution along the $x$-plane, toward the drain contact (fig. \ref{fig grap fpm pxy 1}-b,c), show that a new particle beam around $\mathbf{p}=0$ appears. This second pulse describes the particles coming from the source contact and cumulate along the channel. %
Similar consideration hold for the $\Sigma^-$ particles (we do not report here the $\Sigma^-$ distribution $f^-$), with the difference that in this case the band is almost full and, for small values of the momentum, a large number of particles are able to overcome the small potential gap between the $\Sigma^-$ and the $\Sigma^+$ band (which is equal to $2v_F |\mathbf{p}|$). As a consequence, $\Sigma^-$ particles coming from the source contact, $x=L$, are decelerated by the electric field and instead to be completely reflected back to the source contact, they leave the $\Sigma^-$ band. These particles, now belonging to the $\Sigma^+$ band, are accelerated by the electric field and contribute to the increase of the high energy electron beam depicted in fig. \ref{fig grap fpm pxy 1}-d.
The opposite Klein processes, where the particle flow is directed from the $\Sigma^+$ to the $\Sigma^-$ band is still present but with a smaller intensity and not visible in our plot scale.
In fig. \ref{fig grap fpm 3} we represent the stationary solution for the distribution function $f^+$ in non-intrinsic graphene for an external potential $V_0=0.3$ eV and for $\mu=0.6$ eV. As shown in fig. \ref{fig sat cur vs mu}, for higher values of the chemical potential $\mu$, the semi-classical intraband current quickly increases. In contrast to the intrinsic case, now we observe that particles in the $\Sigma^+$ band populate higher momentum levels and quantum tunneling becomes less significant.

\section{Effective model}\label{sec eff mod}

A direct solution of the system of Eqs. \eqref{mot fpm}-\eqref{mot fi} and its application to electron transport in a graphene sheet demands a high computational effort.
In this section, by investigating the general properties of the solution, we derive some asymptotic limits where an approximated version of the equation of motion applies. The major problem arises from the approximation of the equation of motion for the interband function $f^i$. The diagonal functions $f^\pm$ (which can be considered as a straightforward generalization of the distribution function of electrons in the upper and in the lower part of the Dirac cone) share similar properties with their classical counterparts, and are rather smooth and stable. On the contrary, the interband function $f^i$ shows high frequency oscillation regimes.

In the study of the electric properties of a solid, like the current-voltage $I-V$ characteristic and the conductivity, it is often of primary interest to obtain a correct description of the non-equilibrium stationary state reached by the system in response to an external perturbation field. In the case of the $I-V$ characteristic, the external perturbation is represented by the gradient of the applied potential. The knowledge of the stationary $I-V$ characteristic is crucial for engineering applications of a material and to its integration into a network.
In particular, different approaches should be adopted if a system is characterized by a single time scale according to which all the interesting observables evolve, or if some observables evolve much faster than the others. In the latter case, these variables identify some ``internal dynamics" of a multi-scale process.
In our system, the interband function $f^i$ is a strongly oscillating function and its ``natural" oscillation frequency $\mathcal{A}$ depends on the momentum $\textbf{p}$. This reflects the general principle of quantum mechanics that a wave function containing a superposition of states with different energies, oscillates with a frequency which is proportional to this internal energy difference. In our case, $f^i(\mathbf{r},\mathbf{p})$ describes a mixture of states belonging to the upper and the lower cone. At a given position $\mathbf{r}$, their mean energy difference is equal to $ 2 v_F |\mathbf{p}|$. This appears explicitly in Eq. \eqref{def coef A}. Because of the high value of the Fermi velocity in graphene, this term induces a dynamical evolution of $f^i$ that can be considered to be considerably faster than the other processes induced by the external field (we remark that the identification of the different time scales in which the two-band quantum system evolves, is practically infeasible with the usual definition of many-band Wigner functions given in Eq. \eqref{wigner transf}). Since $f^{\pm}$ describe states with similar energy, in view of Eq. \eqref{fpm unidim}, no ``natural" oscillation frequency is present in the equation for $f^{\pm}$.

\begin{figure}[!t]
\begin{center}
\includegraphics[width=0.450\textwidth,height=0.37\textwidth]{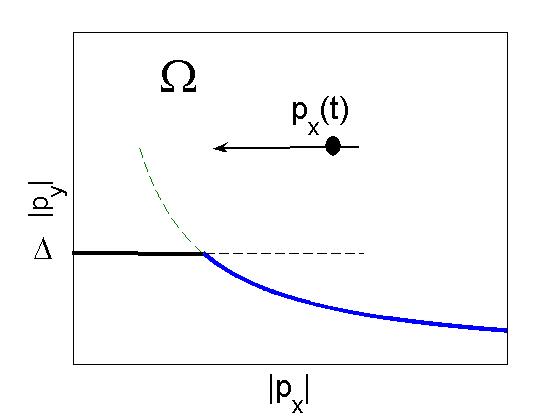}\\%
\caption{Schematic representation of the region $\zO$. 
}\label{fig Omega}
\end{center}
\end{figure}

We are interested in deriving an approximated formula that integrates the function $f_{i}(\mathbf{r},\mathbf{p},t)$. Equation \eqref{fi unidim} can be recast in integral form as
\begin{eqnarray}
f^i (\mathbf{r},\mathbf{p},t )
&=&  \int^{t-t_0}_0 e^{i \int_{0}^{t'} \mathcal{A} ( p_x+\mathcal{E}  \tau ,p_y)   \dif \tau }  \mathcal{D} (p_x+\mathcal{E}  t')  f^d(\mathbf{r}; p_x+\mathcal{E} t',p_y;t- t') \dif t' \; , \label{fi integral 4}
\end{eqnarray}
where $\mathcal{A}$ is given by Eq. \eqref{def Q},  $f^d = f^+-f^-$ and
\begin{eqnarray}
 \mathcal{D}(p_x) &=&   -i  \frac{ \mathcal{E}  }{2  }   \frac{p_y\left( p_x-ip_y\right)}{\left( p_x^2+p_y^2\right)^{3/2}}  \; . \nn 
\end{eqnarray}
After some algebra we obtain
\begin{eqnarray}
f^i (\mathbf{r},\mathbf{p},t ) &=&  -  \frac{ p_y+i  p_x   }{2 \sqrt{ p_y^2+ p_x^2 } }
e^{     i\frac{ \za }{2} \xi(\zb)  } \int^{\zb + \zg  }_{\zb  } \frac{e^{    - i \za \frac{1}{2} \xi(u)}}{ u^2+1}  f^d \left(\mathbf{r}; u,p_y;t- \frac{ p_y u-p_x}{\mathcal{E}}\right)   \dif u \nn \\ \label{fi integral 3}
\end{eqnarray}
with $\za=  \frac{2 v_F p_y^2  }{\hbar\mathcal{E} }  $, $\zb= \frac{p_x}{p_y} $, $\gamma= \frac{\mathcal{E}  (t-t_0)}{p_y}$ and $\xi(\zb)= \zb \sqrt{1+\zb^2}+ \ln ( \zb + \sqrt{1+\zb^2}) $.
We intend to obtain an asymptotic expression for the function $f^i$ in the limit $t\rightarrow \infty$.
A simple analysis of Eq. (\ref{fi integral 3}) reveals that the function $f^i$ displays two qualitatively different behaviors if $p_y$ is grater or smaller  as a certain value $\Delta$. In particular, $f^i$ is smooth if $p_y<\Delta$ and becomes strongly oscillatory otherwise.
In the following, we make this statement more precise. In the hypothesis of smooth $f^d$, the long-time behavior of the function $f^i$ can be estimated by studying the integral
\begin{eqnarray}
\mathcal{I} &=&    \int^{\infty }_{\zb  }
\frac{e^{- i  \frac{\za}{2} \xi(u)}}{ u^2+1}  \dif u \; .
\end{eqnarray}
This expression suggest to estimate $\mathcal{I}$ by means of the stationary phase approximation. This approximation applies when $\za\dot{\xi} (\zb) \gg 1 $ and the exponential is fast oscillating in the scale of the polynomial  decay $(u^2+1)^{-2}$. Explicitly, this condition gives
$|p_y| \gg \Delta \equiv \sqrt{ \frac{\hbar \mathcal{E}}{v_F}} $ or $|p_x| \gg \Delta^2/p_y$. In order to give an analytical estimation of $f^i$, we divide the plane $p_x-p_y$ in the interior and exterior part of the region $\zO$ defined by
\begin{eqnarray}
\zO &=&  \left\{ \mathbf{p}: |p_y| > \Delta \quad or \quad |p_x| > \Delta^2/|p_y| \right\},
\end{eqnarray}
which is depicted in fig. \ref{fig Omega}.
For $\mathbf{p}\in \zO$ the stationary phase approximation applies and the integral of Eq. (\ref{fi integral 3}) can be easily estimated. On the contrary, for $\mathbf{p}\notin \zO$, a different approximation is adopted. Some numerical tests are presented in Appendix \ref{sec num test} where  the validity of the approximation procedure used for the derivation of the asymptotic evolution equation is investigated.

The momentum $p_x$ is evaluated along the trajectory $p_x(t)=p_x(t_0) + \mathcal{E} (t-t_0) $. For $\mathbf{p}\in \zO$ the phase velocity $\mathcal{A}$ changes in time and increases indefinitely for $t$ going to infinity.
The form of Eq. \eqref{fi integral 3} reveals that the long time-behavior of the solution $f^i$ is dominated by the exponential term.
By expanding the function in the exponential up to the third order around the stationary point $u=0$, we obtain the following approximation for $f^i$: \begin{eqnarray}
f^i (\mathbf{r},\mathbf{p},t ) &\simeq&  \pi  \theta(-   \mathcal{E} p_x  ) \textrm{sgn}( p_x p_y ) \frac{ p_y+i  p_x   }{\sqrt{ p_y^2+ p_x^2 } }   \sqrt[3]{\frac{2}{\za}}  \textrm{Ai} \left(\sqrt[3]{ 2 \za^2}\right)  f^d \left(\mathbf{r},\mathbf{p},t \right)  e^{ i\frac{ \za }{2} \xi(\zb) } \; ,\label{app fi py gg}
\end{eqnarray}
where  $\theta$ denotes the Heaviside step function and $\textrm{Ai}(x)$ the Airy function:
\begin{equation}\label{Ai func}
 2\pi  \; \textrm{Ai} \left(x\right)=\int_{-\infty}^\infty e^{i\left(x t+\frac{{t}^3}{3}  \right)}\dif t \; .
\end{equation}
Up to the first order of the electric field, $f^d$ in Eq. \eqref{fi integral 3} can be approximated by $f^{d}_0=f^{+}_0-f^{-}_0$, where $f^{\pm}_0=\left(1+ e^{\pm  v_F|\mathbf{p}|    } \right)^{-1} $ are the Fermi distributions. By expanding  $f^d_0$ around the stationary point $p_x=0$, we obtain
\begin{eqnarray}
\left.f^d_0 (\mathbf{r}; p_x+\mathcal{E} t,p_y;t)\right|_{p_x=0}  \simeq  \frac{t^2}{4} \frac{v_F \mathcal{E}^2 }{|p_y|}  \;   .\label{stima evol fpm}
\end{eqnarray}
For a temperature $T=300$ K, an electric field $\mathcal{E}= 0.1\; \textrm{eV}\mu\textrm{m}^{-1} $ and a parallel momentum $p_y/\hbar= 0.1\; \textrm{nm}^{-1}$ (which are the typical values for graphene), the previous equation reveals that, around the stationary point $p_x=0$, the $f^d_0$ function evolves in a time scale of picoseconds. This time scale is considerably smaller than the ``natural" frequency $\mathcal{A}$, which is of the order of femtoseconds. These considerations suggest to simplify the evolution of the system by assuming that the time evolution of the diagonal functions $f^{\pm}$ is smooth compared to the time evolution of $f^i$. In this hypothesis, we consider an asymptotic model where the function $f^d$ is assumed to be constant around the stationary point $p_x=0$.

For $\mathbf{p}\notin \zO$ (for the sake of simplicity we assume also $|\mathbf{p}|\rightarrow 0$), we approximate the function $\mathcal{A}$ given in Eq. \eqref{def Q} with the dominant contribution
\begin{eqnarray}
\mathcal{A} &\simeq&     - \mathcal{E}  \frac{p_y}{ p_x^2+p_y^2}   \label{def Q app}
\end{eqnarray}
and obtain from Eq. \eqref{fi integral 3} after some algebra
\begin{eqnarray}
 f^i (\mathbf{r},\mathbf{p},t )
&\simeq&  - \frac{1 }{2  } \frac{ p_y+i  p_x   }{\sqrt{ p_y^2+ p_x^2 } } f^d(\mathbf{r},\mathbf{p},t) \left[ \frac{\pi}{2} \textrm{sgn} (\mathcal{E} p_y)
- \tan^{-1} \left(\frac{p_x }{p_y}\right) \right].  \label{app fi py ll}
\end{eqnarray}
This approximations, together with Eq. \eqref{app fi py gg}, lead to the following equations of motion
\begin{eqnarray}
\dpp{ f^\pm }{t}   &=& \pm v_F \frac{p_x}{ \sqrt{ p_x^2+p_y^2} }   \dpp{ f^\pm}{x}+ \mathcal{E}  \dpp{f^\pm}{p_x} \mp \mathcal{T} (f^+-f^-)  \;, \label{fpm app con trans rate}    \\
\mathcal{T}(p_x,p_y)   &=& \left\{
                             \begin{array}{cc}
                                \frac{ \mathcal{E}}{2  } \frac{p_y   }{ p_x^2+p_y^2}  \left[ \frac{\pi}{2} \textrm{sgn} (\mathcal{E} p_y)
- \tan^{-1} \left(\frac{p_x }{p_y}\right) \right] & \textrm{p} \notin  \zO  \;, \\[4mm]
                            -  \pi  \frac{ \mathcal{E} p_y   }{ p_x^2+p_y^2}  \sqrt[3]{\frac{2}{\za}} \textrm{Ai} \left(\sqrt[3]{ 2 \za^2}\right)  \theta(-   \mathcal{E} p_x  ) \textrm{sgn}( p_x p_y )  \Re \left\{   e^{     i\frac{ \za }{2} \xi(\zb) }\right\}      & \textrm{p} \in  \zO \;. \\
                             \end{array}
                           \right. \label{transi rete}
\end{eqnarray}
In contrast to the full quantum mechanical formulation of the dynamics given by Eqs. \eqref{mot fpm}-\eqref{mot fi} or Eqs. \eqref{spin wig vf}-\eqref{spin wig f0}, this form of the  approximated  equations of motion reveals itself to be a simpler and easily understandable description of the interband coherent quantum tunneling phenomena. Here, the transition of a particle between the two bands is modeled by a balance equation, where the transition probability is given in terms of the ``tunneling scattering rate" $\mathcal{T}$. The scattering processes is described in a simple way; in the presence of an electric field $\mathcal{E}$ (directed for simplicity along the direction $x$), the component of the momentum $p_x$ parallel to $\mathcal{E}$ changes according to the Newton law $p_x(t)=p_x(t_0) + \mathcal{E} (t-t_0) $ and when $p_x \simeq 0$ tunneling occurs. Basically, in Eq. (\ref{transi rete}), we distinguish between small ($p_y < \Delta$) and large ($p_y > \Delta$) parallel momenta (we remark that $p_y$ is unaffected by the presence of the electric field and plays the role of a parameter). As explained below in more details, for $p_y < \Delta$ the transition rate $\mathcal{T}$ is proportional to $ 1/|\mathbf{p}|$. Under this condition the band-to-band transition becomes highly favorable and can be interpreted by a quasi-instantaneous process taking place when the particle is at rest. In the opposite limit $p_y > \Delta$, a complex pattern of interference between the two bands appears, which gives rise to the highly oscillatory shape of $f^i$. This oscillatory behavior is captured by the stationary phase approximation, and appears in Eq. \eqref{app fi py gg} through the phase $ \za \xi(\zb) $ which modulates the transition rate $\mathcal{T}$. This part of the solution is the origin of some numerical noise that can be observed in a direct numerical discretization of the equation of motion \eqref{mot fpm}-\eqref{mot fi}. In particular, concerning the application of these equations to a graphene sheet by using a reasonable size of the mesh grid (the number of the grid points for the $\mathbf{p} $ axis being of the order of $10^{2}-10^{3}$), the function $\za \xi(\zb)/(2\pi)$ covers many periods within each $p_x$ discretization cell, making the numerical solution quite inaccurate. This analysis suggests that a possible solution to this problem is to substitute the function $f^i$ by its Gaussian convolution around each discretization point in the $p_x-p_y$ plane. Anyway, despite the uneasy form of $f^i$, the strong oscillation regime prevents any interesting phenomena to emerge at the macroscopic scale (the expectation values of any observable being expressed by the integral of the Wigner functions, so that high oscillating contributions average to zero). Moreover, for $p_y \gg \Delta$ the transition rate $\mathcal{T}$, and consequently the interband tunneling probability, decreases exponentially. This exponential decay agrees with the well known Landau-Zener formula for which the transition probability is proportional to $e^{- \pi \frac{ v_F p_y^2  }{\hbar\mathcal{E} } }$ \cite{Wittig_05}. To go more into details, the formula of Eq. \eqref{transi rete} approaches the Landau-Zener probability in the limit of $|p_y|$ going to infinity. In this case, $\mathcal{T}$ can be simplified by using the asymptotic expression $\textrm{Ai} (x) \simeq \frac{e^{-\left(\frac{2}{3} x^{\frac{3}{2}}\right)}}{2\sqrt{\pi}x^{1/4} }$. We obtain
\begin{eqnarray*}
\mathcal{T}(p_x,p_y)  &\simeq& -    \frac{ \mathcal{E} p_y   }{ p_x^2+p_y^2}  \sqrt{\frac{\pi}{\za}} \frac{      e^{-\left( \frac{4 \sqrt{2}}{3} \frac{  v_F p_y^2  }{\hbar\mathcal{E} }  \right)} }{2^{3/4}   } \theta(-   \mathcal{E} p_x  ) \textrm{sgn}( p_x p_y )  \Re \left\{   e^{     i\frac{ \za }{2}    \xi(\zb) }\right\},
\end{eqnarray*}
where we note that the exponential decay is well represented, but with a slightly different rate (we found $\frac{4 \sqrt{2}}{3} $ instead $\pi$).

The interesting regime for studying the Klein tunneling process is given by $|\mathbf{p}|<\Delta$. From Eq. \eqref{transi rete} we observe the emergency of  some interesting limits revealing deeper insight into the physical description of the tunneling processes provided by our formalism. We consider Eq. \eqref{transi rete} for $p_x=0$. In this case, the transition interband probability $\mathcal{T}$ becomes
\begin{eqnarray*}
\mathcal{T}(0,p_y)   &=& \frac{ |\mathcal{E}|}{ |p_y| }   \frac{\pi}{4} \;,
\end{eqnarray*}
which goes to infinity when $p_y$ approaches zero. We show that this divergence reflects the well know property that a particle, whose trajectory passes exactly through the point in the energy spectrum where the upper and the lower cones touch ($\Gamma$ point), has a unitary probability to pass from one band to the other one. At the $\Gamma$ point, the distinction between the upper and the lower band becomes artificial. For this reason, the distribution function $f^+$ for $\Sigma^+$ particles and $f^-$ for $\Sigma^-$ particles should be equal at $\mathbf{p}=0$ (they represent the same quantity). Any configuration of the system where $f^+(\textbf{r},\textbf{0})\neq f^-(\textbf{r},\textbf{0})$ is unphysical. In our formalism, it is easy to see that the equation of motion ensures automatically that this condition is satisfied at any time. In this context, we can think about our two-band model as a system representing the evolution equation for two populations of strongly interacting particles. The scattering kernel is now written in the relaxation time approximation, where the relaxation time goes to zero. The equation of motion for the difference $f^+-f^-$ is
\begin{eqnarray*}
\dpp{ (f^+ - f^-)}{t}   &=& \pm v_F \frac{p_x}{ \sqrt{ p_x^2+p_y^2} }   \dpp{ (f^+ + f^-)}{x} - 2 \mathcal{T} (f^+-f^-).
\end{eqnarray*}
In the limit $\mathcal{T} \rightarrow \infty$ (Chapman-Enskog limit), we neglect the drift term and the previous equation gives
\begin{eqnarray*}
\dpp{ (f^+ - f^-) }{t}   &=& - 2 \mathcal{T} (f^+-f^-),
\end{eqnarray*}
and up to the order $o(1/\mathcal{T})$ we obtain $f^+=f^-$. A careful analysis of the origin of the divergence in the transition rate $\mathcal{T}$ reveals that the divergent term is exactly the Barry connection associated with our two-band system (see section \ref{sec berry phase}). This can be seen by noting that in the approximation of the phase $\mathcal{A}$ given in Eq. \eqref{def Q app}, we retain the Berry connection $ \frac{1 }{ |\mathbf{p}|^2}   \left(\mathbf{p} \wedge \nabla_\mathbf{r} U   \right)_z$. We discard the contribution $ \frac{ 2 v_F }{ \hbar  } |\mathbf{p}| $ that equals the difference of energy between states with same momentum but localized in different bands. We remark that even in our model a divergent term appears explicitly. This does not lead to an unphysical result that is usually found in similar situations. Our model is able to include explicitly the divergence of the Berry connection by simply forcing the solution $f^d$ to be equal to zero in the region where the Berry phase is not defined.

We consider now more generally the limit $p_y\rightarrow 0 $. Care have to be taken in evaluating this limit. Instead of considering directly the function $\mathcal{T}$, it is convenient to consider the main integral value of this function in an interval $\mathcal{J}=[p_x-\epsilon ,p_x-\epsilon]$ and we let $\epsilon$ go to zero at the end of the calculation:
\begin{eqnarray*}
\overline{\mathcal{T}}(p_x,p_y)   &\equiv&  \frac{1}{2\epsilon} \int_{p_x-\epsilon}^{p_x+\epsilon} \mathcal{T}(p_x,p_y) \dif p_x
= \frac{1}{2\epsilon} \frac{ \mathcal{E}}{2  }  \left[  \frac{\pi}{2} \textrm{sgn} (\mathcal{E} p_y)   \tan^{-1} (v) -\left.   \frac{1}{2}   \left( \tan^{-1}\right)^2 (v) \right|_{(p_x-\epsilon)/p_y}^{(p_x+\epsilon)/p_y}   \right].
\end{eqnarray*}
If $p_x\neq0 $ we choose $\epsilon < |p_x|$ and obtain
\begin{eqnarray*}
\lim_{p_y\rightarrow 0 }\overline{\mathcal{T}}(p_x,p_y)   &=& 0.
\end{eqnarray*}
For $p_x=0$ the limit yields
\begin{eqnarray*}
\epsilon\lim_{p_y\rightarrow 0 }\overline{\mathcal{T}}(p_x,p_y)   &=&  \frac{\pi^2}{8 }    |\mathcal{E}|.
\end{eqnarray*}
These considerations show that the correct limit for the transition probability $\mathcal{T}$ is given by
\begin{eqnarray*}
\lim_{p_y\rightarrow 0 } \mathcal{T}(p_x,p_y)   &=&  \frac{\pi^2}{8 }    |\mathcal{E}| \delta (p_x) \; .
\end{eqnarray*}
This form of the transition rate put in evidence that the band transition around the Dirac point is a strongly localized process.
Based on the previous analysis of the behavior of the transition rate $\mathcal{T}$, we further simplify its expression by evaluating the main value of $\mathcal{T}$. 
We integrate the transition probability with respect to $p_y$,
\begin{eqnarray*}
   \int_{-\Delta}^{\Delta} \mathcal{T}(p_x,p_y) \dif p_y &\simeq&   -  \frac{|\mathcal{E} |}{|p_x|} \frac{\pi}{2}   \left[ 1- \textrm{sgn} (\mathcal{E} p_x) \right]  \log \left[1+ \left(\frac{ \Delta     }{ p_x} \right)^2 \right],
  \end{eqnarray*}
where, for the sake of simplicity, we approximate $\tan^{-1} \left(\frac{p_x }{p_y} \right)\simeq \frac{\pi}{2} \textrm{sgn} \left(p_x p_y \right)$.
We note that $\mathcal{T}$ is non zero only if the momentum of the particle $p_x$  has the opposite sign with respect to the electric field $\mathcal{E}$. According to macroscopical considerations, this means that a particle undergoes a transition only if it is decelerated by the field (the transition takes place when the momentum of the particle can be considered to be small compared to $\Delta$). We remark that this consideration applies irrespective of the cone in which the particle belongs to, since the classical equation of motion for the momentum is $\dot{\mathbf{p}}=\mathcal{E}(\mathbf{x})$ in both cases.

\section{Full quantum solution}\label{sec num sol full quant}

We focus our attention to the numerical solution of the full quantum mechanical electron-hole pair evolution. We consider Eq. \eqref{mot Neum sec} without any further approximation. For the sake of clarity, we report here the equations of motion
\begin{eqnarray}
i \hbar \dpp{ \mathcal{S}'  }{t}   &=&   \left[\mathcal{H}'(\mathbf{r},\mathbf{p}) , \mathcal{S}' \right]_\star \; ,\label{mot Neum2}
\end{eqnarray}
where
\begin{eqnarray}
\mathcal{H}'&=&\mathcal{U}'\left(\mathbf{r},\mathbf{p}\right) + \Lambda(\mathbf{p})\; ,  \\[2mm]
\mathcal{U}'\left(\mathbf{r},\mathbf{p}\right)  &=&  \frac{1}{\left(2\pi\right)^2 }\int   {\Theta}  \left(  \mathbf{p}+\frac{ \hbar}{2}  \boldsymbol{\mu}  \right)    {\Theta}^\dag   \left(  \mathbf{p} - \frac{ \hbar}{2} \boldsymbol{\mu}    \right)  U (\mathbf{r}'  ) e^{i (\mathbf{r}-\mathbf{r}')\cdot \boldsymbol{\mu} }   \dif  \boldsymbol{\mu} \dif \mathbf{r}' \; ,  \label{rela fra sym2}
\end{eqnarray}
and
\begin{eqnarray}
\left[\mathcal{H}' , \mathcal{S}' \right]_\star
&=&\frac{1}{\left(2\pi\right)^{4} }  \int \left[\mathcal{H}' \left( \mathbf{r}-\frac{ \hbar}{2}   \boldsymbol{\zh} ,\mathbf{p} +\frac{ \hbar}{2} \boldsymbol{\mu} \right)   \mathcal{S}' \left( \mathbf{r}' ,\mathbf{p}' \right)  - \mathcal{S}' \left( \mathbf{r}' ,\mathbf{p}' \right)  \mathcal{H}'\left( \mathbf{r}+\frac{ \hbar}{2}   \boldsymbol{\zh} ,\mathbf{p} -\frac{ \hbar}{2} \boldsymbol{\mu} \right)  \right] \nn \\ && \hspace{4cm }\times \; e^{i (\mathbf{r}-\mathbf{r}')\cdot \boldsymbol{\mu} + i (\mathbf{p}-\mathbf{p}')\cdot \boldsymbol{\zh}}   \dif  \boldsymbol{\mu} \dif \mathbf{r}'  \dif  \boldsymbol{\zh} \dif \mathbf{p}'  \; . \nn
\end{eqnarray}
In order to obtain a numerically tractable model, instead to solve the full $4$-dimensional system (two-dimension both in position and in momentum), we consider a simpler case where the solution is uniform along the $y$ direction (but non-constant with respect the momentum along the same direction) and we solve the reduced system in $\mathbb{R}^3$. In this hypothesis, we have
\begin{eqnarray*}
   \mathcal{S}' \left( \mathbf{r},\mathbf{p}\right)& = &\mathcal{S}' \left(r_x ;p_x,p_y\right)  \; , \\
   \mathcal{U}'\left( \mathbf{r},\mathbf{p}\right) &= & \mathcal{U}'\left( r_x;  p_x,p_y\right)
\end{eqnarray*}
and the equation of motion simplifies to
\begin{eqnarray*}
\left[\mathcal{U}' , \mathcal{S}' \right]_\star
&=& \frac{1}{\left(2\pi\right)^{2} }  \int \left[\mathcal{U}' \left( r -\frac{ \hbar}{2}  \zh  , p  +\frac{ \hbar}{2} \mu , p_y \right)   \widetilde{ \mathcal{S}'} ( \mu , \zh ,p_y )  - \widetilde{ \mathcal{S}'} ( \mu , \zh ,p_y )  \mathcal{U}'\left( r +\frac{ \hbar}{2}  \zh , p - \frac{ \hbar}{2} \mu  , p_y \right)  \right] \\ && \hspace{0.7\textwidth} \times e^{i   r    \mu  + i  p  \zh  }   \dif \mu   \dif \zh  \; ,   \nn\\
\widetilde{ \mathcal{S}'} ( \mu , \zh  ,p_y) &=&  \int  \mathcal{S}' \left(  r_x',p_x',p_y \right)    e^{-i  ( r_x'  \mu  +   p'_x \zh ) }    \dif r_x'    \dif p_x'  \nn
\end{eqnarray*}
with
\begin{eqnarray*}
\mathcal{U}'(r ,p,p_y )
& =&   \frac{1}{\left(2\pi\right)  }\int   {\Theta}  \left( p+\frac{ \hbar}{2}  \mu  , p_y \right)    {\Theta}^\dag   \left(p - \frac{ \hbar}{2} \mu  ,p_y  \right)   \int  U ( r'  ) e^{  i(r- r') \mu   }   \dif r'   \dif   \mu   \; ,
\end{eqnarray*}
where we put in evidence that in this case the coordinate $p_y$ plays the role of a parameter. An analysis of these expressions reveals that in order to obtain an efficient numerical scheme, it is convenient to impose
\begin{eqnarray*}
\Delta_r= \za \Delta_\eta/2 \; ,\\
\Delta_p= \zb \Delta_\mu/2  \; ,
\end{eqnarray*}
where $\za,\zb$ are integers (or inverse of integer) and $\Delta_z$ denotes the size of the numerical discretization of the $z$ axes. Since the $x-\mu$ and $p-\eta$ are conjugate variables, the discrete Fourier transform (DFT) requires (see for example \cite{Fensley_90})
\begin{eqnarray*}
\Delta_r= \frac{2\pi}{N_r \Delta_\mu }\; , \\
\Delta_p= \frac{2\pi}{N_p \Delta_\eta }
\end{eqnarray*}
and we obtain that the following relationship should be fulfilled:
\begin{eqnarray*}
\frac{N_r}{N_p} = \frac{\zb}{ \za } \; .
\end{eqnarray*}
Finally, we consider the following first-order (in time) solution of Eq. \eqref{mot Neum2} that shows itself to be particular stable and weakly affected by numerical noise:
\begin{eqnarray*}
\mathcal{S}' \left(r,p;t+\Delta_t\right)
&=& \frac{1}{\left(2\pi\right)^{2} }  \int e^{-\frac{i\Delta_t}{\hbar}\mathcal{U}' \left( r -\frac{ \hbar}{2}  \zh , p  +\frac{ \hbar}{2} \mu  \right)}   \widetilde{ \mathcal{S}'} ( \mu , \zh,t  )  e^{ \frac{i\Delta_t}{\hbar}\mathcal{U}' \left( r +\frac{ \hbar}{2}  \zh , p  -\frac{ \hbar}{2} \mu  \right)}      e^{i   r    \mu  + i  p  \zh  }   \dif \mu   \dif \zh    \nn,
\end{eqnarray*}
where the matrix $e^{ i \Delta \mathcal{U}' \left( r, p \right)} $ is evaluated by the formula
\begin{eqnarray*}
e^{ i \Delta \mathcal{U}' \left( r, p \right)} &=&\left(\cos(|\mathbf{u}|) u_0 + i \sin(|\mathbf{u}|) \frac{\mathbf{u}\cdot \bs{\sigma} }{|\mathbf{u}|} \right) e^{i u_0} \; , \\
\mathbf{u}&=& \frac{\Delta}{2} \; \textrm{tr} \left\{\mathcal{U}' \left( r, p \right) \cdot \bs{\sigma}\right\}\; ,\\
u_0&=& \frac{\Delta}{2} \; \textrm{tr} \left\{\mathcal{U}' \left( r, p \right) \right\}
\end{eqnarray*}
with tr denoting the trace of the $2\times2$ matrix.
\begin{figure}[!t]
\begin{center}
\framebox{\begin{minipage}[h]{0.48\textwidth}
\includegraphics[width=\textwidth,height=0.74\textwidth]{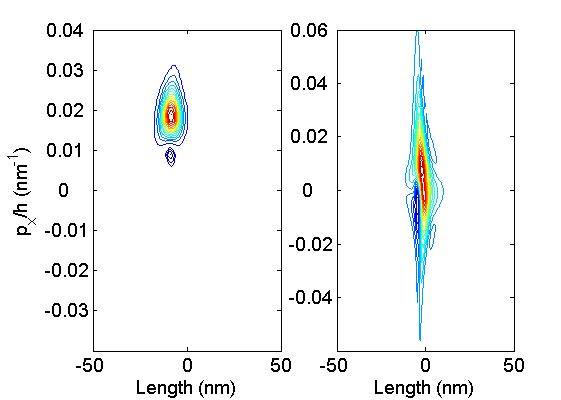} \\
\flushleft  \vspace{-12mm} a)
\end{minipage}}
\framebox{\begin{minipage}[h]{0.48\textwidth}
\includegraphics[width=\textwidth,height=0.74\textwidth]{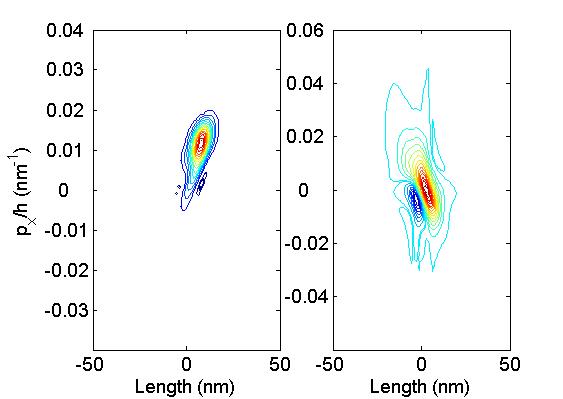} \\
 \flushleft \vspace{-12mm} b)
\end{minipage}}\\
\framebox{\begin{minipage}[h]{0.48\textwidth}
\includegraphics[width=\textwidth,height=0.74\textwidth]{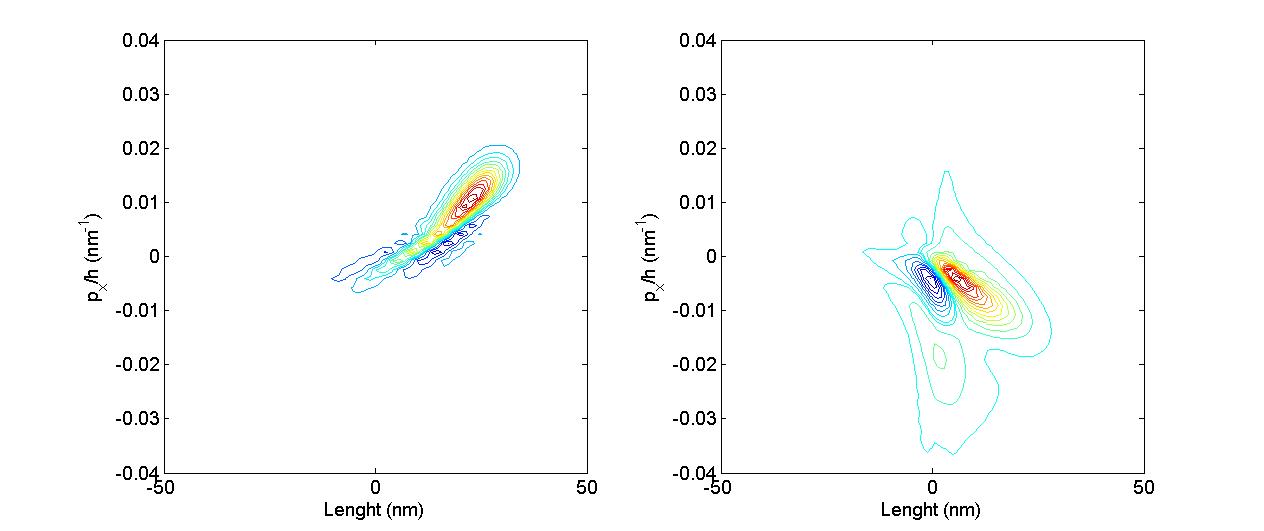} \\
\flushleft  \vspace{-12mm} c)
\end{minipage}}
\framebox{\begin{minipage}[h]{0.48\textwidth}
\includegraphics[width=\textwidth,height=0.74\textwidth]{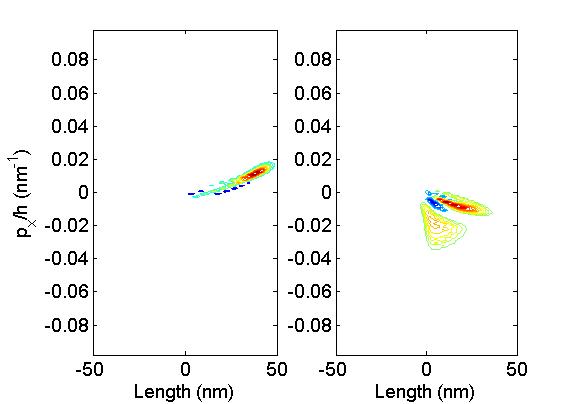} \\
 \flushleft \vspace{-12mm} d)
\end{minipage}}
\caption{Contour plot of $f^+$ (left plot) and $f^-$ (right plot) for a) $t= 300 $ fs; b) $t= 500 $ fs; c) $t= 700 $ fs; d) $t= 900 $ fs. Here $p_y/\hbar=10^{-2}$ nm$^{-1}$ and $V_0=0.1$ eV.
}\label{fig cont f_pm_1}
\end{center}
\end{figure}

\begin{figure}[!t]
\begin{center}
\framebox{\begin{minipage}[h]{0.48\textwidth}
\includegraphics[width=\textwidth,height=0.74\textwidth]{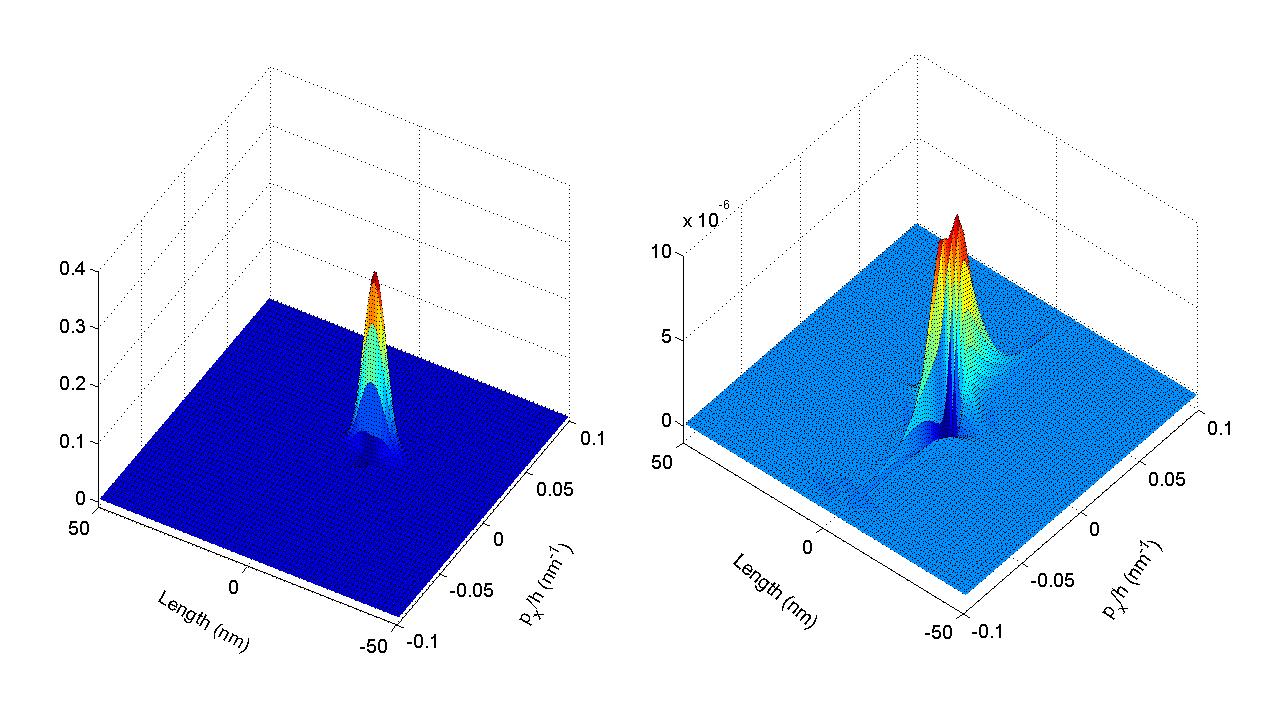} \\
\flushleft  \vspace{-12mm} a)
\end{minipage}}
\framebox{\begin{minipage}[h]{0.48\textwidth}
\includegraphics[width=\textwidth,height=0.74\textwidth]{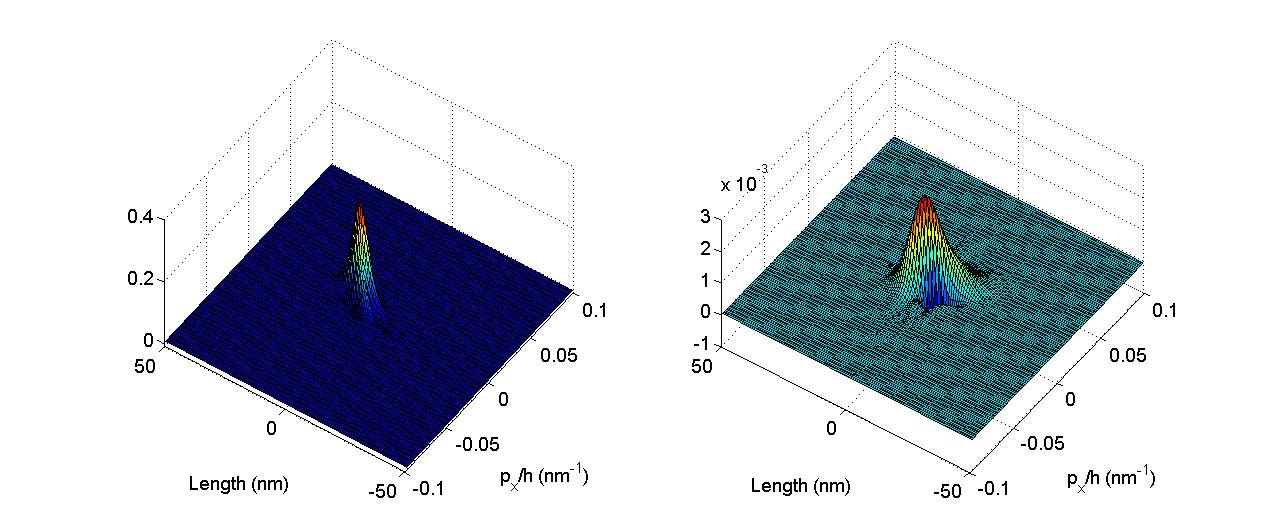} \\
 \flushleft \vspace{-12mm} b)
\end{minipage}}\\
\framebox{\begin{minipage}[h]{0.48\textwidth}
\includegraphics[width=\textwidth,height=0.74\textwidth]{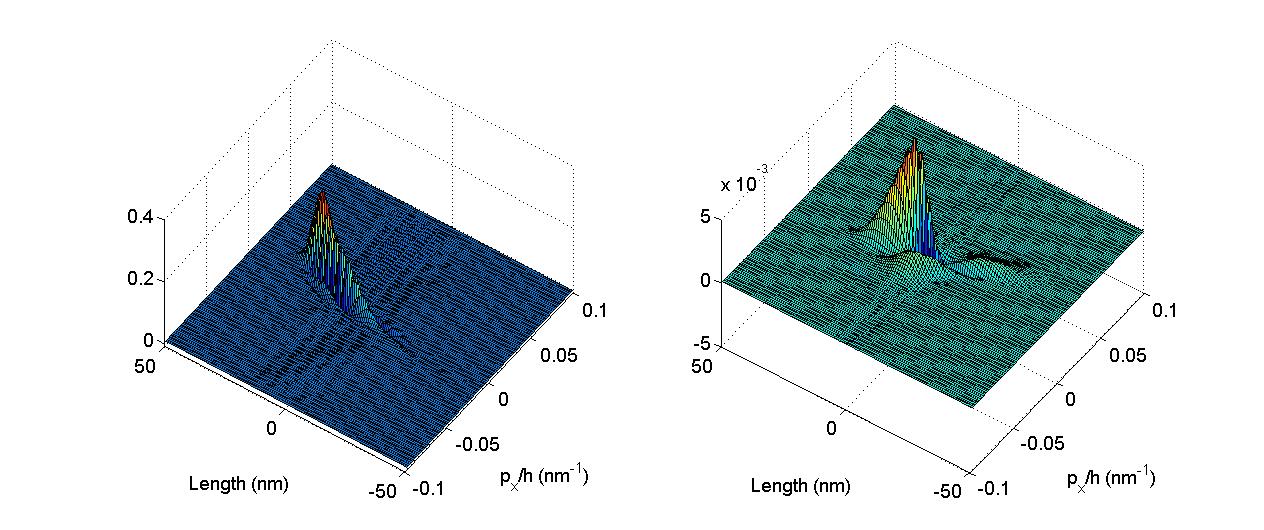} \\
\flushleft  \vspace{-12mm} c)
\end{minipage}}
\framebox{\begin{minipage}[h]{0.48\textwidth}
\includegraphics[width=\textwidth,height=0.74\textwidth]{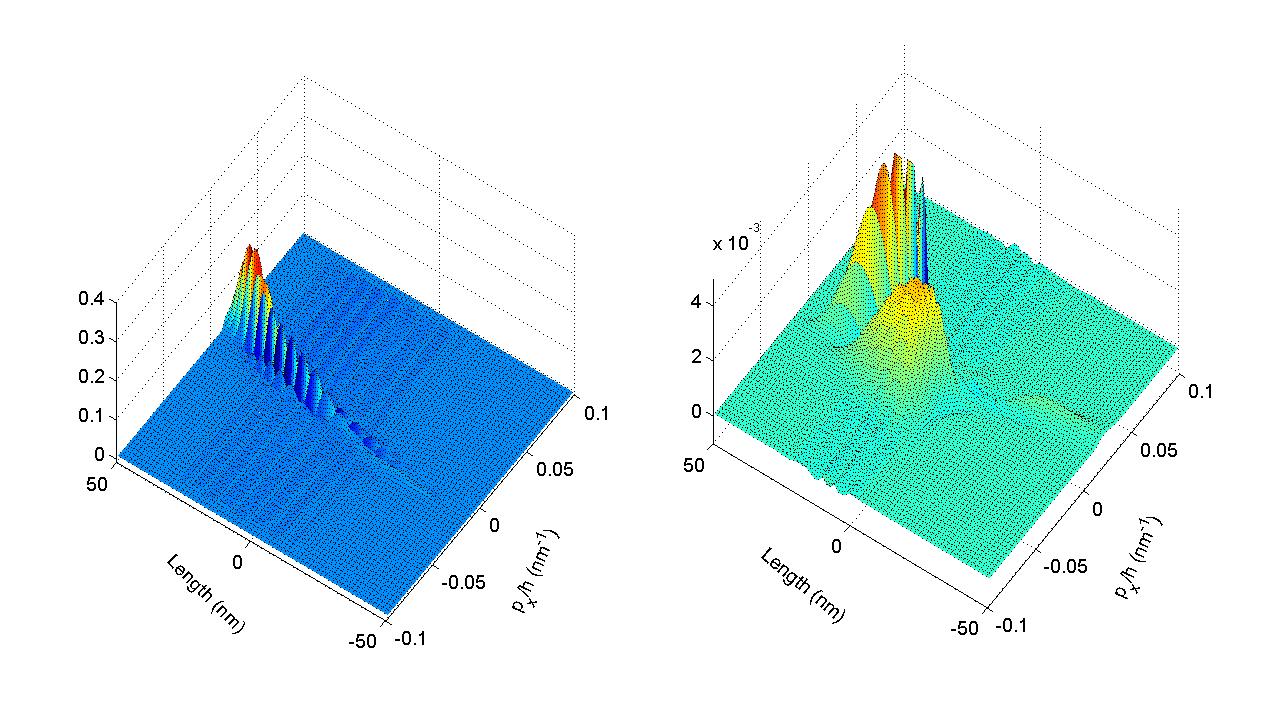}
 \flushleft \vspace{-12mm} d)
\end{minipage}}
\caption{3D plot of $f^+$ (left plot) and $f^-$ (right plot) for a) $t= 300 $ fs; b) $t= 500 $ fs; c) $t= 700  $ fs; d) $t= 900 $ fs. Here $p_y/\hbar=10^{-2}$ nm$^{-1}$ and  $V_0=0.1$ eV. 
}\label{fig 3D f_pm_1}
\end{center}
\end{figure}

\begin{figure}[!t]
\begin{center}
\framebox{\begin{minipage}[h]{0.48\textwidth}
\includegraphics[width=\textwidth,height=0.74\textwidth]{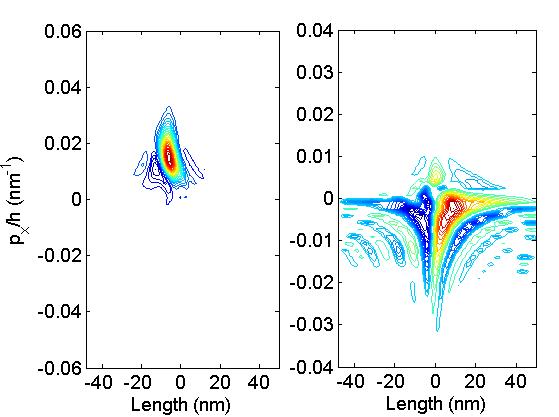} \\
\flushleft  \vspace{-12mm} a)
\end{minipage}}
\framebox{\begin{minipage}[h]{0.48\textwidth}
\includegraphics[width=\textwidth,height=0.74\textwidth]{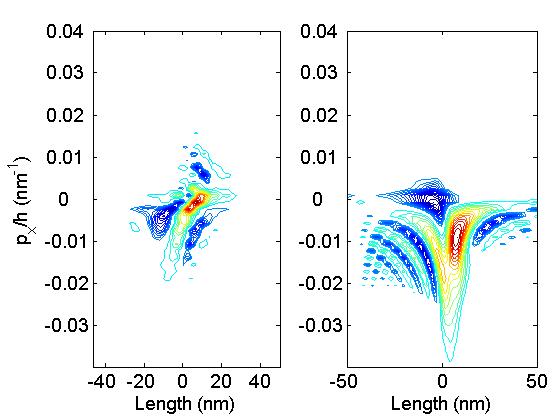} \\
 \flushleft \vspace{-12mm} b)
\end{minipage}}\\
\framebox{\begin{minipage}[h]{0.48\textwidth}
\includegraphics[width=\textwidth,height=0.74\textwidth]{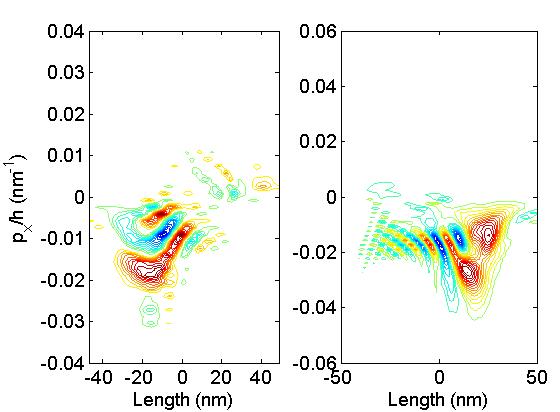} \\
\flushleft  \vspace{-12mm} c)
\end{minipage}}
\framebox{\begin{minipage}[h]{0.48\textwidth}
\includegraphics[width=\textwidth,height=0.74\textwidth]{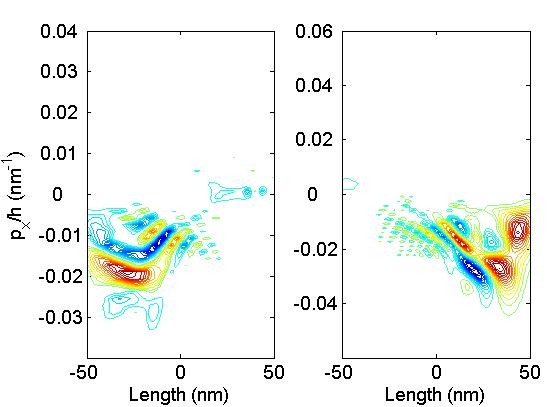}
 \flushleft \vspace{-12mm} d)
\end{minipage}}
\caption{Contour plot of $f^+$ (left plot) and $f^-$ (right plot) for a) $t= 300 $ fs; b) $t= 500 $ fs; c) $t= 700 $ fs; d) $t= 900 $ fs. Here $p_y/\hbar=10^{-4}$ nm$^{-1}$ and  $V_0=0.03$ eV. 
}\label{fig cont f_pm_2}
\end{center}
\end{figure}

\begin{figure}[!t]
\begin{center}
\framebox{\begin{minipage}[h]{0.48\textwidth}
\includegraphics[width=\textwidth,height=0.74\textwidth]{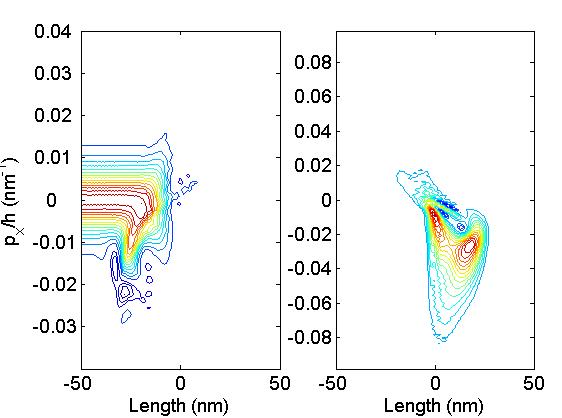} \\
\flushleft  \vspace{-12mm} a)
\end{minipage}}
\framebox{\begin{minipage}[h]{0.48\textwidth}
\includegraphics[width=\textwidth,height=0.74\textwidth]{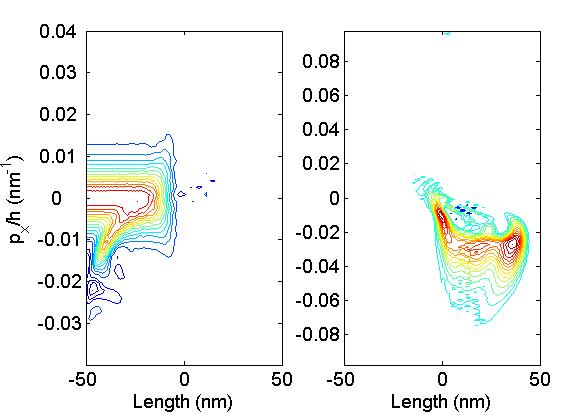} \\
 \flushleft \vspace{-12mm} b)
\end{minipage}}\\
\framebox{\begin{minipage}[h]{0.48\textwidth}
\includegraphics[width=\textwidth,height=0.74\textwidth]{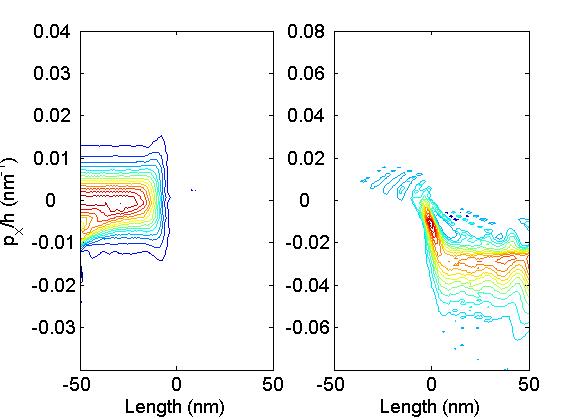} \\
\flushleft  \vspace{-12mm} c)
\end{minipage}}
\framebox{\begin{minipage}[h]{0.48\textwidth}
\includegraphics[width=\textwidth,height=0.74\textwidth]{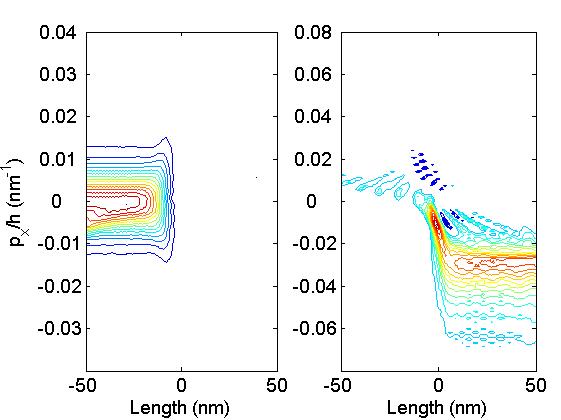}
 \flushleft \vspace{-12mm} d)
\end{minipage}}
\caption{Contour plot of $f^+$ (left plot) and $f^-$ (right plot) for a) $t=300$ fs; b) $t=500$ fs; c) $t=700$ fs; d) $t=900$ fs. Here $p_y/\hbar=10^{-2}$ nm$^{-1}$ and $V_0=0.03$ eV.
}\label{fig cont f_pm_3}
\end{center}
\end{figure}

\begin{figure}[!t]
\begin{center}
\framebox{\begin{minipage}[h]{0.48\textwidth}
\includegraphics[width=\textwidth,height=0.74\textwidth]{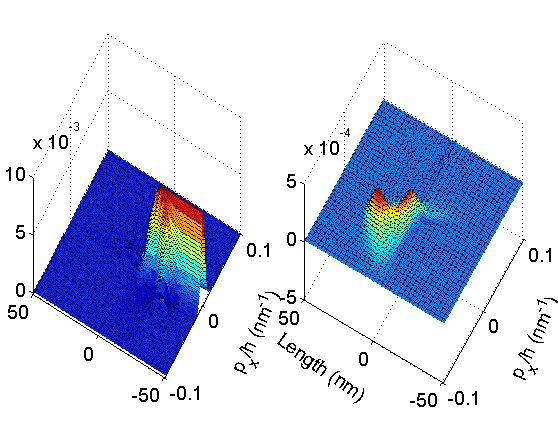} \\
\flushleft  \vspace{-12mm} a)
\end{minipage}}
\framebox{\begin{minipage}[h]{0.48\textwidth}
\includegraphics[width=\textwidth,height=0.74\textwidth]{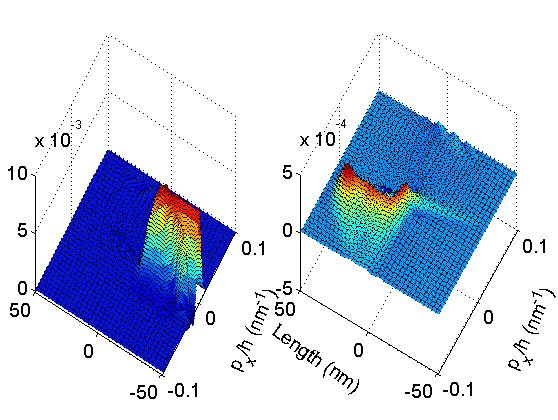} \\
 \flushleft \vspace{-12mm} b)
\end{minipage}}\\
\framebox{\begin{minipage}[h]{0.48\textwidth}
\includegraphics[width=\textwidth,height=0.74\textwidth]{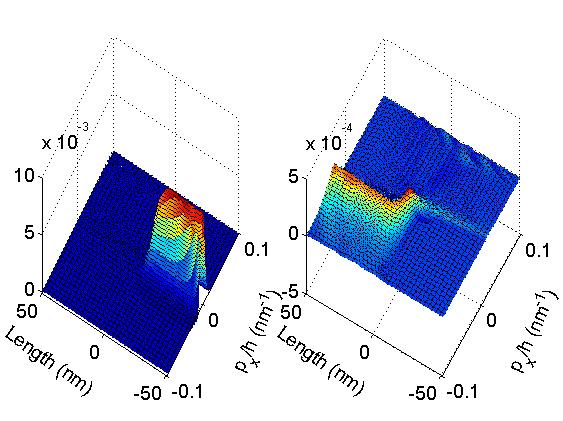} \\
\flushleft  \vspace{-12mm} c)
\end{minipage}}
\framebox{\begin{minipage}[h]{0.48\textwidth}
\includegraphics[width=\textwidth,height=0.74\textwidth]{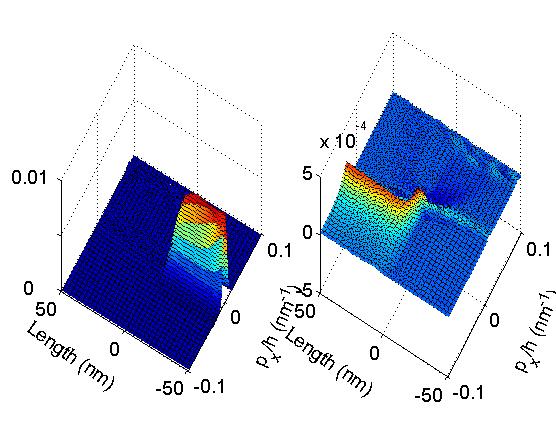}
 \flushleft \vspace{-12mm} d)
\end{minipage}}
\caption{3D plot of $f^+$ (left plot) and $f^-$ (right plot) for a) $t=300$ fs; b) $t=500$ fs; c) $t=700$ fs; d) $t=900$ fs. Here $p_y/\hbar=10^{-2}$ nm$^{-1}$ and $V_0=0.3$ eV.
}\label{fig 3D f_pm_3}
\end{center}
\end{figure}
In order to present the structure and to give a general impression of the form of the full quantum mechanical solution based on the matrix Wigner function $\mathcal{S}'$, we consider a standard text-book case, in which a minimum uncertainty Gaussian packet impacts a potential barrier.
The study is depicted in fig. \ref{fig cont f_pm_1}. In particular, we consider as initial condition a Gaussian pulse in the upper graphene cone ($\Sigma^+$ band) localized around the position $x_0=-40\; \textrm{nm}$ and momentum $p_0/\hbar=10^{-2}\; \textrm{nm}^{-1}$ and with a parallel momentum $p_y/\hbar= 10^{-2}  \; \textrm{nm}^{-1}$. Furthermore, we assume a vanishing initial condition for the $\Sigma^-$ band and a vanishing band-to-band correlation (represented by the function $f^i$). The shape of potential barrier used in the simulation is depicted in fig. \ref{fig pseudo_pot}, but here we consider a lower barrier of $0.1$ eV. The wave is initially localized in the zero potential region ($x<0$) and is directed against the potential step. The height of the barrier is chosen in order to be smaller than the mean kinetic energy of the wave packet, so that it could be overcomed by the Gaussian packet. In fig. \ref{fig cont f_pm_1} we display the solution for different times (from sub-panel $a$ to $d$). In particular, we show the contour plot of $f^+$ (on the left side) and $f^-$ (on the right side). The solution shows that the $\Sigma^+$ packet overcomes, as expected, the potential barrier but also generates a transmitted particle beam in the $\Sigma^-$ band. We note that the $f^-$ function is initially generated very close to the potential barrier, where the main momentum of the $\Sigma^+$ particles tends to the minimum. In order to give an impression of the relative width of the two solutions, in fig. \ref{fig 3D f_pm_1} we depict the 3D version of fig. \ref{fig cont f_pm_1}.

In the second numerical test (depicted in fig. \ref{fig cont f_pm_2}), we consider a more stressed case consisting of a higher barrier (difference of potential equal to $0.3$ eV) and a lower parallel momentum $p_y/\hbar=10^{-4} \; \textrm{nm}^{-1}$. According to the previous discussion, when the momentum of the particles approaches the Dirac point $\mathbf{p}=0$, the coupling between the functions $f^+$ and $f^-$ increases considerably. The emergence of a divergence in the coupling terms entails that the numerical solution becomes more and more critical. This consideration explains the emergency of the complex interference pattern that is observed in the phase plane where several ripples appear both in the upper and in the lower cone distributions. Anyway, the main classical features of the solution are preserved (especially in the classical-like region $|x|>50$ nm). We see that the incoming particles impact the potential barrier and are reflected back. Besides, a transmitted pulse in the lower cone is generated. We note that both pulses stay ``mainly" positive with some residual oscillation induced by the band-to-band interference.

Finally, we address our attention to a more realistic case, where an ohmic contact is localized on the left part of the domain at $x=-50 $ nm. We model the contacts in the usual way by assuming an incoming thermal equilibrium distribution for the $f^+$ particles (to highlight the Klein phenomenon, we artificially impose vanishing boundary condition for the $f^-$ distribution). We put evidence into the effect of the full quantum band-to-band tunneling (and also to study transient effects) by initially discarding any band-to-band effects. For this purpose, we take as initial condition for $f^+$ the full quantum \emph{single band} thermal equilibrium distribution. For $t>0$ we  allow particles to pass from one band to the other one by solving the complete two-band system. We observe that, after a transient regime, a constant flux of particles is generated in the $\Sigma^-$ band. This new particle beam can be interpreted in the classical language as a flux of particles initially created around $p_x=0$ and subsequently accelerated by the electric field (that is a barrier for the particles in the upper cone and an accelerating field for those localized in the lower cone) and propagate afterwards freely in the classical ($x>50$ nm) zone. In particular, we remark that the solution shows the nice property that the transmitted particle beam stays manly positive (at least within the numerical precision of our simulation). Here, the hight of the barrier is equal to $0.3$ eV and $p_y/\hbar=10^{-2} \; \textrm{nm}^{-1}$. In fig. \ref{fig 3D f_pm_3} we depict the 3D profile of the solution.

\section{Conclusion}

In this contribution, the ballistic transport of electrons in graphene by including quantum effects is investigated in terms of the Wigner formalism. The resulting formulation reveals itself to be particularly close to the classical description of the particle motion. Special attention is devoted to model the Klein tunneling and to study the correction to the total current in intrinsic graphene induced by this phenomenon. Due to the high numerical complexity of the resulting system of equations, an approximated closed-form solution is obtained.
The simulations show that for an intrinsic graphene in the presence of a strong electric field, our model predicts a non-negligible correction to the charge inside the device. Some numerical experiments are performed where the evolution of a Gaussian pulse in the presence of a potential barrier is investigated. The numerical solutions show that our formalism shares some nice properties with the classical solution like smoothness and positivity in the regions sufficiently faraway form the potential barrier.

\section{Appendix}

\subsection{Derivation of Eq. \eqref{mot Neum sec}}\label{app for H Hp}

In this section, we derive the equation of motion \eqref{mot Neum sec}. We recall some properties of the Weyl operator algebra.
Coherently with the notation used in sec. \ref{sec mod}, we will denote by $\widehat{\mathcal{A}} = \mathcal{W}\left[\mathcal{A}\right] $ the operator associated with the phase-space function $\mathcal{A}\left(\mathbf{r},\mathbf{p}\right)$. The following property holds true
\begin{eqnarray}
\mathcal{W}^{-1} \left[\widehat{\mathcal{A}}\; \widehat{\mathcal{B}} \right]  &=& \widehat{\mathcal{A}} \star \widehat{\mathcal{B}}. \label{prod moy}
\end{eqnarray}
In the hypothesis that $\mathcal{A} $ and  $\mathcal{B}$ are sufficiently smooth, the Moyal product defined in Eq. \eqref{star prod} admits the following $\hbar$-expansion:
\begin{eqnarray}
\mathcal{A}\star \mathcal{B} &=& \sum_n \left(\frac{i\hbar}{2}\right)^n \frac{1}{n!}\mathcal{A} (\mathbf{r},\mathbf{p}) \left[ \overleftarrow{\nabla_\mathbf{r}}\cdot \overrightarrow{\nabla_\mathbf{p}}- \overleftarrow{\nabla_\mathbf{p}} \cdot\overrightarrow{\nabla_\mathbf{r}}  \right]^n \mathcal{B}(\mathbf{r},\mathbf{p}) \label{moy pro 3}\\  &=&
\sum_n\sum_{k=0}^n   \left(\frac{i\hbar}{2}\right)^n \frac{(- 1)^k }{n!} {{n }\choose{k }} \mathcal{A} (\mathbf{r},\mathbf{p}) \left( \overleftarrow{\nabla_\mathbf{r}} \cdot\overrightarrow{\nabla_\mathbf{p}} \right)^{n-k}  \left( \overleftarrow{\nabla_\mathbf{p}} \cdot \overrightarrow{\nabla_\mathbf{r}}  \right)^k \mathcal{B}(\mathbf{r},\mathbf{p}), \label{moy pro 2}
\end{eqnarray}
where the arrows indicate on which operator the gradients act. In particular, if both operators depend only on one variable ($\mathbf{r}$ or $\mathbf{p}$), the Moyal product becomes the ordinary product
\begin{eqnarray}
\mathcal{A} (\mathbf{p})\star \mathcal{B}  (\mathbf{p}) =\mathcal{A} (\mathbf{p})  \mathcal{B}  (\mathbf{p}). \label{simpl moy prod}
\end{eqnarray}
The Moyal product can be expressed also in integral form:
\begin{eqnarray}
\mathcal{A}\star \mathcal{B}
&=&\frac{1}{\left(2\pi\right)^{4} } \mathcal{A} \left( \mathbf{r} ,\mathbf{p} \right)  \; e^{\frac{i\hbar}{2} \left(\overleftarrow{\nabla_\mathbf{r}} \overrightarrow{\nabla_\mathbf{p}}- \overleftarrow{\nabla_\mathbf{p}} \overrightarrow{\nabla_\mathbf{r}}\right) }  \int  \mathcal{B} \left( \mathbf{r}' ,\mathbf{p}' \right)   e^{i (\mathbf{r}-\mathbf{r}')\cdot \boldsymbol{\mu} + i (\mathbf{p}-\mathbf{p}')\cdot \boldsymbol{\zh}}   \dif  \boldsymbol{\mu} \dif \mathbf{r}'  \dif  \boldsymbol{\zh} \dif \mathbf{p}' \nn \\
&=&\frac{1}{\left(2\pi\right)^{4} }  \int \mathcal{A} \left( \mathbf{r}-\frac{ \hbar}{2}   \boldsymbol{\zh} ,\mathbf{p} +\frac{ \hbar}{2} \boldsymbol{\mu} \right)    \mathcal{B} \left( \mathbf{r}' ,\mathbf{p}' \right)   e^{i (\mathbf{r}-\mathbf{r}')\cdot \boldsymbol{\mu} + i (\mathbf{p}-\mathbf{p}')\cdot \boldsymbol{\zh}}   \dif  \boldsymbol{\mu} \dif \mathbf{r}'  \dif  \boldsymbol{\zh} \dif \mathbf{p}' \;,  \label{prod moy expl 2}
\end{eqnarray}
where we used the expansion of Eq. \eqref{moy pro 3} in the expression
\begin{eqnarray*}
\mathcal{A} \left( \mathbf{r} ,\mathbf{p} \right)  \; e^{\frac{i\hbar}{2} \left(\overleftarrow{\nabla_\mathbf{r}} \overrightarrow{\nabla_\mathbf{p}}- \overleftarrow{\nabla_\mathbf{p}} \overrightarrow{\nabla_\mathbf{r}}\right) }     e^{i (\mathbf{r}-\mathbf{r}')\cdot \boldsymbol{\mu} + i (\mathbf{p}-\mathbf{p}')\cdot \boldsymbol{\zh}}  =
  \mathcal{A} \left( \mathbf{r} ,\mathbf{p} \right)  \; e^{-\frac{ \hbar}{2} \left(\overleftarrow{\nabla_\mathbf{r}}  \boldsymbol{\zh}- \overleftarrow{\nabla_\mathbf{p}}   \boldsymbol{\mu} \right) }     e^{i (\mathbf{r}-\mathbf{r}')\cdot \boldsymbol{\mu} + i (\mathbf{p}-\mathbf{p}')\cdot \boldsymbol{\zh}} \; .
\end{eqnarray*}
In same way we obtain
\begin{eqnarray}
\mathcal{B} \star \mathcal{A}
&=&\frac{1}{\left(2\pi\right)^{4} }  \int    \mathcal{B} \left( \mathbf{r}' ,\mathbf{p}' \right)  \mathcal{A} \left( \mathbf{r}+\frac{ \hbar}{2}   \boldsymbol{\zh} ,\mathbf{p} -\frac{ \hbar}{2} \boldsymbol{\mu} \right)  e^{i (\mathbf{r}-\mathbf{r}')\cdot \boldsymbol{\mu} + i (\mathbf{p}-\mathbf{p}')\cdot \boldsymbol{\zh}}   \dif  \boldsymbol{\mu} \dif \mathbf{r}'  \dif  \boldsymbol{\zh} \dif \mathbf{p}' \; . \label{prod moy expl 3}
\end{eqnarray}
We evaluate the Moyal symbol $\mathcal{H}'$ of the transformed Hamiltonian $\widehat{\mathcal{H}}'=\widehat{\Theta}\widehat{\mathcal{H}}\widehat{\Theta}^\dag$. By using Eq. \eqref{prod moy} we obtain
\begin{eqnarray}
\mathcal{H}'\left(\mathbf{r},\mathbf{p}\right)  &=& \Theta \star      {\mathcal{H}}  \star \widetilde{\Theta^\dag}, \label{Hp app}
\end{eqnarray}
where $\widetilde{\Theta^\dag} \equiv \mathcal{W}^{-1} \left[\widehat{\Theta}^\dag  \right]$.
In particular,  $\widetilde{\Theta^\dag} = \left( \mathcal{W}^{-1} \left[ \widehat{\Theta}\right]\right)^\dag = \Theta^\dag $. This can be verified easily by applying the Weyl operator to the relationship $\widehat{\Theta} \widehat{\Theta}^\dag = \mathcal{I}$, where $\mathcal{I} $ denotes the identity operator. We obtain
\begin{eqnarray}
  \mathcal{I} =  \Theta \star \widetilde{\Theta^\dag} =  \Theta \; \widetilde{\Theta^\dag}=\Theta \;  {\Theta^\dag},
\end{eqnarray}
where in the second equality, we used the expansion of Eq. \eqref{moy pro 3}. The symbol $\Theta$ defined in Eq. \eqref{theta symb} does not depend on the spatial variable $\mathbf{r}$. Eq. \eqref{Hp app} thus becomes
\begin{eqnarray}
\mathcal{H}'\left(\mathbf{r},\mathbf{p}\right)  &=&  \Theta \star \left[\mathcal{H}_0(\mathbf{p}) + U(\mathbf{r}) \right] \star \Theta^\dag
 =\Lambda(\mathbf{p}) + \Theta \star  U(\mathbf{r}) \star \Theta^\dag,    \label{Hp app2}
\end{eqnarray}
where we used Eq. \eqref{simpl moy prod} and Eq. \eqref{diag H}. Proceeding as in Eq. \eqref{prod moy expl 2} we obtain
\begin{eqnarray*}
\mathcal{U}' &\equiv& {\Theta} ( \mathbf{p}) \star  {U} ( \mathbf{r})  \star {\Theta}^\dag ( \mathbf{p}) =  {\Theta} ( \mathbf{p})e^{-\frac{i\hbar}{2} \left(\overleftarrow{\nabla_\mathbf{p}} \overrightarrow{\nabla_\mathbf{r}} \right) }  {U} ( \mathbf{r})  e^{ \frac{i\hbar}{2} \left(\overleftarrow{\nabla_\mathbf{r}} \overrightarrow{\nabla_\mathbf{p}} \right) } {\Theta}^\dag ( \mathbf{p}) \\
& =&  \frac{1}{\left(2\pi\right)^2 }\int   {\Theta} ( \mathbf{p})e^{\frac{ \hbar}{2} \left(\overleftarrow{\nabla_\mathbf{p}}  \boldsymbol{\mu}  \right) }   U (\mathbf{r}'  ) e^{i (\mathbf{r}-\mathbf{r}')\cdot \boldsymbol{\mu} }e^{ - \frac{ \hbar}{2} \left(  \boldsymbol{\mu}   \overrightarrow{\nabla_\mathbf{p}} \right) } {\Theta}^\dag ( \mathbf{p})\dif  \boldsymbol{\mu} \dif \mathbf{r}'\\
& =&  \frac{1}{\left(2\pi\right)^2 }\int   {\Theta}  \left(  \mathbf{p}+\frac{ \hbar}{2}  \boldsymbol{\mu}  \right)    {\Theta}^\dag   \left(  \mathbf{p} - \frac{ \hbar}{2} \boldsymbol{\mu}    \right)  U (\mathbf{r}'  ) e^{i (\mathbf{r}-\mathbf{r}')\cdot \boldsymbol{\mu} }   \dif  \boldsymbol{\mu} \dif \mathbf{r}',
\end{eqnarray*}
where we applied the identity
\begin{eqnarray*}
{U} ( \mathbf{r})
&=& \frac{1}{\left(2\pi\right)^2 } \int_{\mathbb{R}_{\zh}^d} \int_{\mathbb{R}_{\mu}^d}  {U} (\mathbf{r}'  ) e^{i (\mathbf{r}-\mathbf{r}')\cdot \boldsymbol{\mu} }\dif  \boldsymbol{\mu} \dif \mathbf{r}' \; .
\end{eqnarray*}

\subsection{Numerical study of the asymptotic model}\label{sec num test}
\begin{figure}[!h]
\begin{center}
 \includegraphics[width=0.450\textwidth,height=0.37\textwidth]{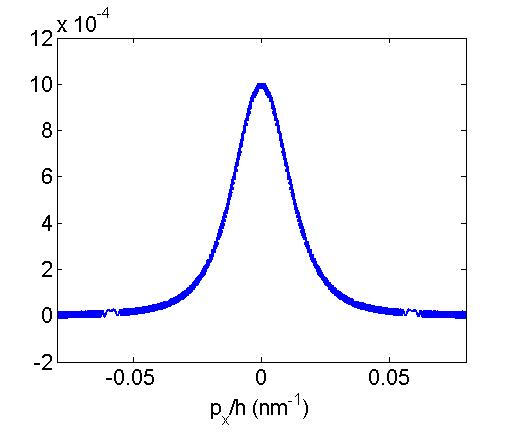}%
 \caption{Stationary values of the function $f^i$ in the presence of a uniform electric field $\mathcal{E}$ for $p_y/\hbar =0.2  \textrm{nm}^{-1}$.}\label{fig fwi_study1}
\end{center}
\end{figure}

\begin{figure}
\begin{center}
\framebox{a)\includegraphics[width=0.450\textwidth,height=0.37\textwidth]{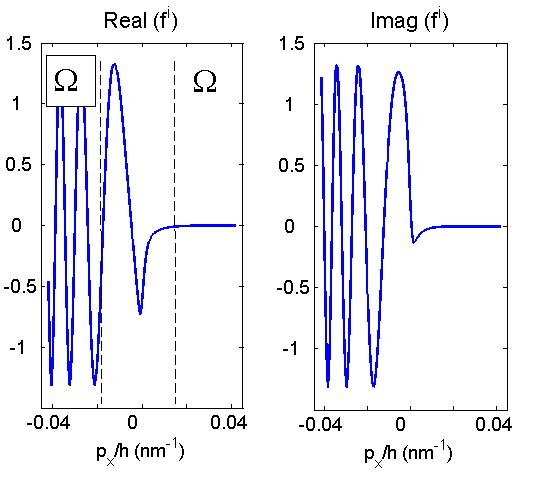}}
\framebox{b)\includegraphics[width=0.450\textwidth,height=0.37\textwidth]{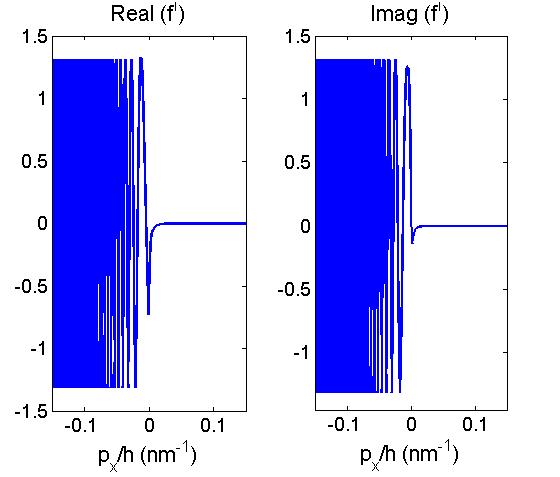}}\\
\framebox{c)\includegraphics[width=0.450\textwidth,height=0.37\textwidth]{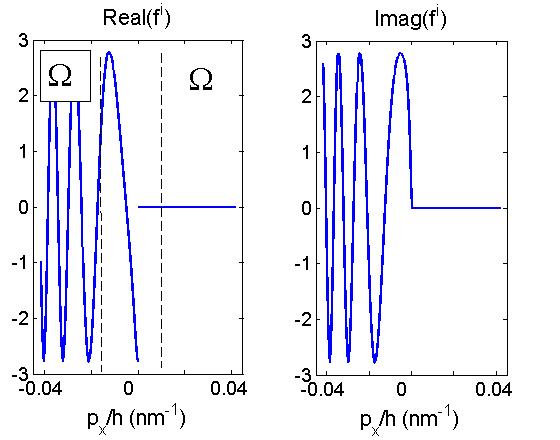}}
\framebox{d)\includegraphics[width=0.450\textwidth,height=0.37\textwidth]{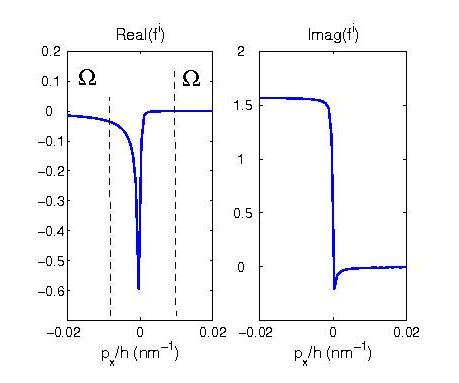}}\\
\caption{a-b) Numerical solution of $f^i$ for $p_y/\hbar = 2 \cdot 10^{-3} \textrm{nm}^{-1}$. c-d) Approximated expression of $f^i$ given in  Eqs. (\ref{app fi py gg}) and (\ref{app fi py ll}).}\label{fig fwi_study3}
\end{center}
\end{figure}

We present here some numerical tests that validate the approximations used in sec. \ref{sec eff mod} and show some characteristic features displayed by the function $f^i$ in correspondence to the different limits previously discussed. We solve Eqs. \eqref{fi integral 4} with high numerical precision and obtain the stationary solution $f^i$ in the presence of a uniform electric field $\mathcal{E} $. We assume for simplicity $f^d=1$.
The numerical results show that, as expected, the function $f^i$ displays high-frequency oscillations along the $p_x$ axis. This behavior becomes more and more evident when the parallel momentum $p_y$ goes to zero. In graphene, the band-to-band transition probability approaches one for $\mathbf{p}=0$. For this reason, small values of $p_y$ characterize the interesting regime, when we study quantum corrections to the interband  current.
Moreover, Eq. \eqref{def Q} shows that for $p_y$ going to zero, the function $f^i$ oscillates with a period of $\Delta_{p_x} \propto \frac{\hbar \mathcal{E}}{v_F p_x}$. The monotonic increase of the oscillation frequency along the $p_x$ axis for increasing values of $p_x$  (we recall that the electric field is directed along the $x$ axis and we are evaluating the integral along the trajectory $p_x(t)=p_x(t_0) + \mathcal{E} (t-t_0) $) makes the direct numerical approximation of $f^i$ quite delicate. Besides, the high oscillating behavior of $f^i$ contrasts the form of the diagonal functions $f^\pm$ that stays smooth even in the presence of significant band-to-band transitions. Thus, it ensures the validity of the stationary phase approximation used in sec. \ref{sec eff mod}.

In our simulations $\mathcal{E} = 0.3 \cdot 10^{-3}\, \textrm{eV}\; \textrm{nm}^{-1}$ and $\Delta= \sqrt{ \frac{\hbar \mathcal{E}}{v_F}}= 2\cdot 10^{-2}\, \textrm{nm}^{-1}$.
Figure \ref{fig fwi_study1} shows the typical form of the solution when $p\in \zO$ ($p_y= 0.2 \; \textrm{nm}^{-1}$).
The function is highly oscillating with a period of nearly $10^{-6}\; \textrm{nm}^{-1} $ which is considerably small with respect the typical spatial variation of the classical distribution functions. To make a comparison, the functions $f^{\pm}$ in graphene at the temperature of 300 K vary on a resolution scale of the order of $10^{-3}\,\textrm{nm}^{-1} $ (see for example fig. \ref{fig grap fpm pxy 1}). According to Eq. \eqref{app fi py gg}, the numerical solution confirms the exponential decay of $f^i$  for increasing values of $p_y$. In fig. \ref{fig fwi_study3} we represent the solution for $p_y/\hbar = 2 \cdot 10^{-3} \textrm{nm}^{-1}$. In particular, in fig. \ref{fig fwi_study3}-a-b we display the function $f^i$ evaluated numerically (figure \ref{fig fwi_study3}-a is the snapshot of the zoom of fig. \ref{fig fwi_study3}-b in the central region) and in  fig. \ref{fig fwi_study3}-c-d we depict the approximation of $f^i$ obtained by Eqs. (\ref{app fi py gg})-(\ref{app fi py ll}). A glance to fig. \ref{fig Omega} shows that if $p_y<\Delta$ (as in the present case), the small values of $p_x$ does not belong to $\Omega$ and should be approximated by Eq. \eqref{app fi py ll}. On the contrary, for increasing values of $p_x$, Eq. \eqref{app fi py gg} applies. For the sake of clearness, we marked in  fig. \ref{fig fwi_study3} the boundary of the region $\Omega$. We remark that around $p_x =0 $, where $p_x\simeq p_y$ (and thus $\mathbf{p}\notin\zO$) the expression of $f^i$ is well reproduced by a simple pole (Eq. \eqref{app fi py ll}). For increasing values of $p_x$ the function starts to oscillate in same way as described by Eq. (\ref{app fi py gg}).



\end{document}